\documentclass[11pt]{article}
\usepackage{jcappub}
\usepackage{epstopdf}
\usepackage{graphicx}
\usepackage{color}
\usepackage[usenames,dvipsnames]{xcolor}
\usepackage{hyperref}
\usepackage[export]{adjustbox}

\usepackage{amsmath}
\usepackage{amssymb}
\usepackage{graphicx}
\usepackage{microtype}
\usepackage{pdfpages}
\usepackage{natbib}
\usepackage{graphicx}
\usepackage{xcolor}
\usepackage{amssymb,amsmath,bm}
\usepackage{hyperref}
\usepackage[export]{adjustbox}
 \usepackage[utf8]{inputenc}
\usepackage{multirow}

\newcommand{\avg}[1]{\ensuremath{\left\langle \,#1\, \right\rangle}}

\newcommand{\be}{\begin{equation}}
\newcommand{\ee}{\end{equation}}

\newcommand{\bea}{\begin{eqnarray}}
\newcommand{\eea}{\end{eqnarray}}
\newcommand{\bdm}{\begin{displaymath}}
\newcommand{\edm}{\end{displaymath}}

\newcommand{\fnu}{$f_\nu$}

\def\Mpc{\, {\rm Mpc}/h}

\def\kMpc{\, h / {\rm Mpc}}

\def\fnu{f_{\nu}}
\def\Pcc{P_{cc}(k)}

\newcommand{\eq}[1]{Eq.~(\ref{#1})}
\newcommand{\eqs}[2]{Eqs.~(\ref{#1},\ref{#2})}

\def\ie{{\em i.e.}~}
\def\eg{{{\em e.g.}~}}

\title{Weighing neutrinos with the halo environment}

\author[a,b,c,1]{Arka Banerjee,\note{Corresponding author.}}
\author[d,e]{Emanuele Castorina,}
\author[f]{Francisco Villaescusa-Navarro,}
\author[g,h]{Travis Court,}
\author[i,j,k,l]{Matteo Viel}

\affiliation[a]{Kavli Institute for Particle Astrophysics and Cosmology, Stanford University, 452 Lomita Mall, Stanford, CA 94305, USA}
\affiliation[b]{Department of Physics, Stanford University, 382 Via Pueblo Mall, Stanford, CA 94305, USA}
\affiliation[c]{SLAC National Accelerator Laboratory, 2575 Sand Hill Road, Menlo Park, CA  94025, USA}
\affiliation[d]{Department of Physics, University of California, Berkeley, CA 94720}
\affiliation[e]{Berkeley Center for Cosmological Physics, Berkeley, CA 94720}
\affiliation[f]{Center for Computational Astrophysics, Flatiron Institute, 162 5th Avenue, 10010, New York, NY, USA}
\affiliation[g]{Allegheny College, 520 N. Main Street, Meadville, PA 16335, USA}
\affiliation[h]{Department of Physics and Astronomy, University of Pittsburgh, 3941 O'Hara Street, Pittsburgh, PA 15260, USA}
\affiliation[i]{SISSA, Via Bonomea 265, 34136 Trieste, Italy}
\affiliation[j]{INFN, Sez. di Trieste, Via Valerio 2, 34127 Trieste, Italy}
\affiliation[k]{IFPU, Institute for Fundamental Physics of the Universe, via Beirut 2, 34151 Trieste, Italy}
\affiliation[l]{INAF,  Osservatorio Astronomico di Trieste, via Tiepolo 11, I-34131 Trieste, Italy}

\emailAdd{arkab@stanford.edu}
\emailAdd{ecastorina@berkeley.edu}
\emailAdd{fvillaescusa@flatironinstitute.org}
\emailAdd{TAC136@pitt.edu}
\emailAdd{matteoviel@gmail.com}

\abstract{Nonlinear objects like halos and voids exhibit a scale-dependent bias on linear scales in massive neutrino cosmologies. The shape of this scale-dependent bias is a unique signature of the neutrino masses, but the amplitude of the signal is generally small, of the order of $f_\nu$, the contribution of neutrinos to the total matter content ($\lesssim 1\%$). In this paper, we demonstrate for the first time how the strength of this signal can be substantially enhanced  
by using information about the halo environment at a range of scales. This enhancement is achieved by using certain combinations of the large scale Cold Dark Matter and total matter environments of halos, both of which are measurable from galaxy clustering and weak lensing  surveys.}

\begin{document}
\maketitle
\section{Introduction and motivations}
The Large Scale Structure (LSS) of the Universe encodes information about all the processes and the cosmological parameters that, throughout the history of the Universe, have shaped the distribution of galaxies we observe today.
Amongst other things, LSS observables are affected by the presence of massive neutrinos \cite{Hu1997,Lesgourgues:2006nd}. Terrestrial oscillation experiments have proved that neutrinos are massive \cite{K2K2003,SuperK98,SNO2001}, but are yet to pin down either the exact mass scale of neutrinos, or the hierarchy of the mass eigenstates. In recent years, cosmological probes \citep{Aghanim:2018eyx,Yeche:2017upn,Palanque-Delabrouille:2014jca} have put some of the strongest constraints on the sum of neutrino masses $M_\nu=\sum m_\nu$. As future surveys \citep{DESI,Euclid,LSST,S4ScienceBook,SIMONS}, both from the ground and space, push both to larger survey volumes and ability to probe smaller scales, 
the current bounds on the total neutrino mass from cosmology is expected to improve dramatically, especially by using combinations of the data from different surveys \cite{Font-Ribera:2013rwa,Schmittfull:2017ffw,Banerjee:2016suz,Brinckmann:2018owf,Yu:2018tem}. Many different studies have pointed out that even if the total neutrino mass, $M_\nu$  is the minimum allowed by oscillation data, i.e. $M_\nu \sim 0.06\,$eV, it would be detectable at a level of roughly $3\sigma$, even in the most conservative survey predictions.

Most of the constraints and forecasts mentioned above attempt to infer the total neutrino mass by measuring an overall damping in the power spectrum of perturbations of Cold Dark Matter (CDM), as well as in the perturbations of all matter, which includes the neutrino component. On small scales, the predicted size of this effect from linear theory is $\sim 8 \fnu$ on the total matter power spectrum, and $\sim 6\fnu$ on the CDM power spectrum, where $\fnu$ is the fraction of matter made up of neutrinos: $f_\nu=\Omega_\nu/\Omega_{\rm m}$. For realistic neutrino masses, $\fnu \lesssim 1\%$, so the absolute size of this damping effect is small, but measurable when the signal-to-noise is high. However, an overall damping of the power spectrum on small scales is degenerate with other cosmological parameters (e.g. $\sigma_8$ \citep{FVN18}), and one has to be careful to account for these degeneracies in the analysis.
This fact has spurred work on alternative ways to constraint neutrino masses beyond the traditional two point analysis \cite{Coulton2018,Li2018,Ruggeri2018,ZhuCastorina}.

Rather than a constant suppression of the power spectrum at small scales, a more unique imprint of massive neutrinos lies in the actual shape of the damping of the power spectrum - i.e. how neutrinos change the  amplitude of the power spectrum $P(k)$ as a function of the wavenumber $k$, over the whole range of $k$ that are accessible to various surveys. This is set by the shape of the transfer function of neutrinos, which, in turn, is set by their masses, and is not degenerate with any of the other standard cosmological parameters. However, the size of this effect is roughly $\sim \fnu$, and is usually difficult to measure. Apart from the small amplitude of this effect, it is worth pointing out that this effect can only be measured on very large scales, where there are usually very few independent modes. 

Related to this, various papers have recently explored whether nonlinear objects like galaxies, halos and voids display a scale-dependent bias on large scales in massive neutrino cosmologies, and how to self-consistently account for this in the analysis of LSS data \cite{Villaescusa-Navarro:2013pva,Castorina:2013wga,LoVerde:2014pxa,Castorina:2015bma,Banerjee:2016zaa,Chiang:2017vuk,Schmittfull:2017ffw,Chiang:2018laa,Munoz:2018ajr,Vagnozzi:2018pwo}. It was recently shown that for dark matter halos, the linear scale-dependent bias, when the bias is measured with respect to the underlying CDM\footnote{On the scales of interest of this study, CDM and baryon density fields track each other closely, so we will refer to the combined CDM and baryon field simply as the CDM field throughout this paper.} field is very small \cite{Chiang:2018laa}, but there is significantly stronger scale dependence when measured with respect to the total matter field \cite{Villaescusa-Navarro:2013pva,Castorina:2013wga,Castorina:2015bma, Raccanelli_19, Valcin_19}. For voids, one of the most sensitive objects to neutrino masses \citep{Paco_13,Massara_15,Kreisch_18,Pisani_19,Cora_19}, selected using CDM only, a similar behavior is reproduced, but for voids defined in the total matter field, a stronger scale dependence in the bias was seen \cite{Banerjee:2016zaa}. In all of these cases, the departure of the linear bias from a scale independent value was shown to be set by the shape of the neutrino transfer function, i.e., a direct function of the neutrino mass for a fixed decoupling temperature.

Having established the fact that measuring a linear scale-dependent bias can be used to robustly infer the neutrino mass, it is useful to explore ways in which the scale-dependent behavior of the bias can be enhanced. To this end, we explore a completely new direction - to use information from the large scale environment of dark matter halos.
It has been shown in a number of studies that the halo environment is an extremely good indicator of the large scale clustering \cite{Shi:2017rwh,Paranjape:2017zpc,Paranjape18b,Alam:2018nxn, Han2018,Salcedo2018,Ramakrishnan19}.
In our particular case, we use the halo environment to define certain population of halos which can exhibit extremely strong scale-dependent bias on large scales, only in massive neutrino cosmologies. The scope of this paper is to present a new method to enhance the size of, and measure, the scale-dependent bias of populations of halos in massive neutrino cosmologies by using information about their CDM and total matter environment. Critically, both of these large scale environments are potentially measurable in galaxy redshift and weak lensing surveys, and recent studies have explored the constraints that can be placed on cosmological parameters using the combination of the two \cite{Gruen:2017xjj,Friedrich:2017qfe,Font-Ribera:2013rwa}. 
We consider our method exploratory, and as such we do not attempt, in this paper, to forecast neutrino mass constraints using this technique in future galaxy surveys where additional observational effects could play an important role.

This paper is organized as follows. In Section 2, we explore the origins of the scale-dependent effects of massive neutrinos in more detail. Section 3 presents the analytical motivation behind our new technique using standard tools for Gaussian random fields.
Section 4 presents the actual measurements of scale-dependent bias of halos sorted by their large scale environments in full N-body simulations including massive neutrinos. Finally, in Section 5, we summarize our findings and discuss the prospects of applying this method for measuring neutrino masses in future surveys.

\section{Scale-dependent effects from massive neutrinos}
\label{sec:scale_dep}
To understand how neutrinos imprint information about their masses on Large Scale Structure, we must first consider the thermal history of the Universe. Neutrinos decoupled as ultra-relativistic particles when the background temperature was $\sim 1\,$MeV, and then freely propagated in the Universe. At the CMB epoch, neutrinos were still relativistic, and contributed to the radiation content, but at late times, when they become non-relativistic, neutrinos contribute to the total matter content. In a Universe with massive neutrinos the total matter density is the sum of the CDM (c), baryons (b), and neutrino components,
\begin{align}
\rho_m = \rho_{c}+\rho_b+\rho{_\nu}\, .
\end{align}
One can also write down the perturbations of field of the total matter content in terms of the perturbations in the individual species:
\begin{align}
\delta_m = f_c \delta_{c} + f_b \delta_b + \fnu \delta_\nu \, ,
\end{align}
where $f_i$, (with $i \in \{c,b,\nu\}$), are the fractional background energy densities of each species, i.e., $f_c=\rho_{c}/\rho_m$,  $f_b=\rho_{b}/\rho_m$ and $f_\nu=\rho_{\nu}/\rho_m$. For the purpose of this work the perturbations in CDM and baryons can be considered identical: $\delta_b=\delta_{c}$ and $f_c+f_b=1-\fnu$. The energy fraction of neutrinos is given by $\fnu=M_\nu/(\Omega_m93.14h^2)$ where $h$ is the dimensionless Hubble parameter, ($h=H_0/100~{\rm kms^{-1}Mpc^{-1}}$). For realistic values of neutrino masses $\fnu \lesssim 0.01$, so the neutrino effects are usually quite small.

Despite being non-relativistic at late times, massive neutrinos still travel much faster than standard cold dark matter particles as their characteristic velocities are set by the redshifted Fermi-Dirac distribution that was fixed at decoupling.
In a Hubble time we can define a characteristic scale associated with the distance traveled by neutrinos, the so called free-streaming length $\lambda_{\text{fs}}$ \citep{Lesgourgues:2006nd}
\begin{align}
\lambda_{\text{fs}}(m_\nu,z) = 7.7 \frac{1+z}{[\Omega_\Lambda+\Omega_m(1+z)^3]^{1/2}}\left(\frac{1\,\text{eV}}{m_\nu}\right)\,\text{Mpc}/h \, ,
\end{align} 
which at low redshift and for light neutrinos can be of $\mathcal{O}(100)\,\Mpc$. Thus, on small scales, the transfer function of neutrinos is very different from that of CDM and baryons, since neutrino clustering will be highly suppressed on scales below $\lambda_{\text{fs}}$. 
\begin{figure}
\centering
\includegraphics[width=0.6\textwidth]{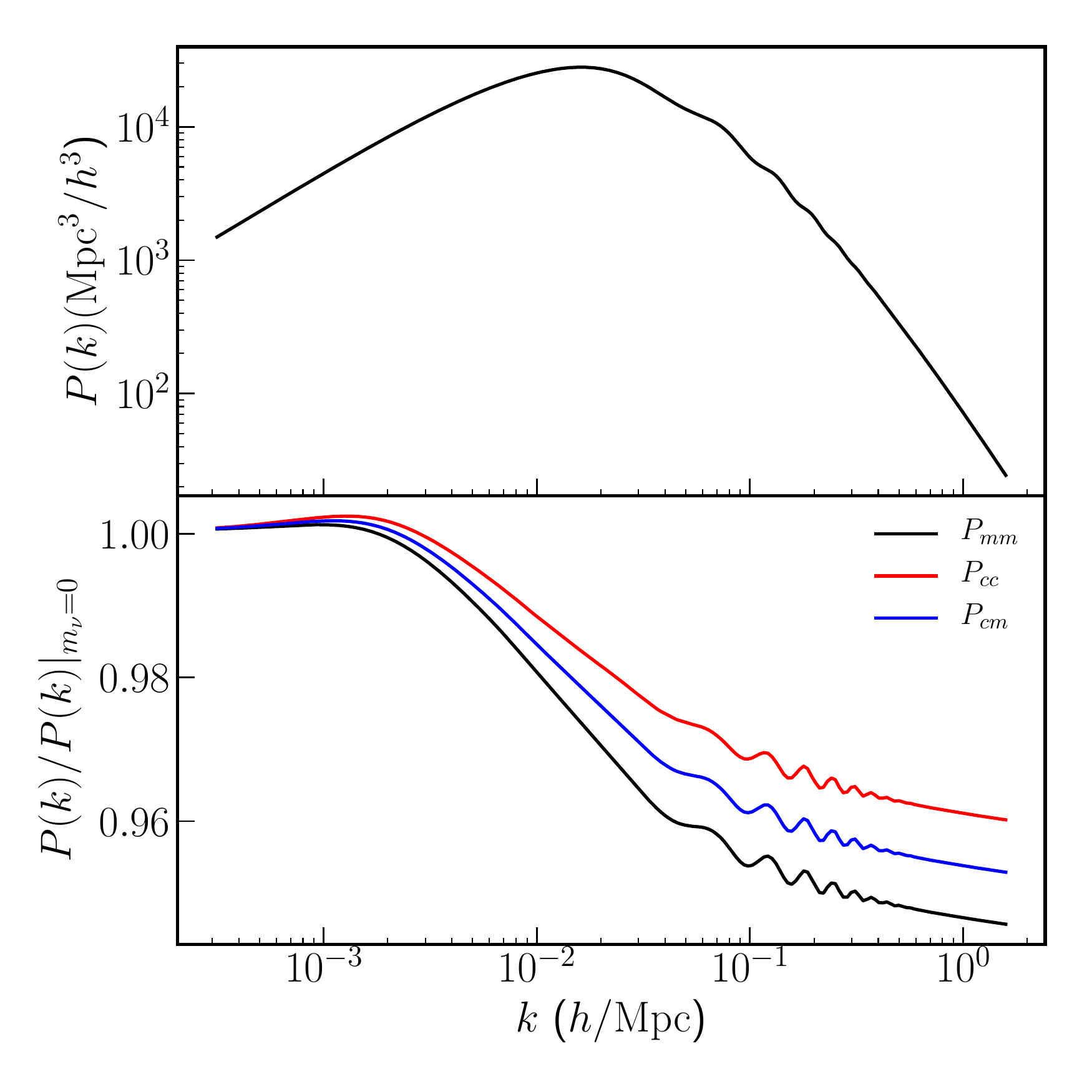}
\caption{\emph {Top panel:} Matter power spectrum at $z=0$ for $M_\nu = 0.09\,$eV with three degenerate neutrino masses. \emph{Bottom panel:} The relative damping of the total matter power spectrum, CDM power spectrum and the matter-CDM power spectrum compared to the massless neutrino cosmology. The damping amplitude on small scales in the matter power spectrum is $\sim 8 \fnu$, while the damping on small scales in the CDM power spectrum is $\sim 6 \fnu$.}
\label{fig:pk}
\end{figure}
Information about the neutrino masses is therefore, directly encoded into the matter power spectrum,
\begin{equation}
\label{eq:Pmk}
P_m(k)  = (1-\fnu)^2P_{cc}(k)+2 \fnu(1-\fnu)P_{c\nu}(k)+\fnu^2 P_{\nu \nu}^2(k) \, ,
\end{equation}
where $P_{cc}(k)$, $P_{\nu\nu}(k)$ are the CDM and neutrino auto-power spectra, respectively, and $P_{c\nu}(k)$ is the CDM-neutrino cross-power spectrum. As can be seen, information about neutrinos explicitly enters in two ways in the above expression: 1) through a change in the background value of $\fnu$, and 2) through the terms that encode neutrino perturbations - $P_{c\nu}$ and $P_{\nu\nu}$. Neutrinos also affect the evolution of CDM perturbations themselves by suppressing, compared to a massless case, the growth of CDM structures on scales smaller than the free streaming scale. In linear theory it can show that at $z=0$ \citep{Lesgourgues:2006nd}
\begin{align}
 \frac{P_{cc}(k;f_\nu)}{P_{cc}(k;f_\nu = 0)} \xrightarrow[]{k\lambda_{\text{fs}}\gg1}1-6 f_{\nu}\,.
\label{eq:cdm_damping}
\end{align}
On very large scales, $k\lambda_{\text{fs}}\ll1$, neutrinos behave like CDM and the three power spectra on the right hand side are identical, whilst on small scales, $k\lambda_{\text{fs}}\gg1$ , neutrino fluctuations are washed out,
\begin{align}
 P_{mm}(k) = \begin{cases}
           P_{cc}(k) & \text{if } k\ll\lambda_{\text{fs}}^{-1} \\
           (1-f_{\nu})^2\,P_{cc}(k) & \text{if } k\gg \lambda_{\text{fs}}^{-1}\,.
          \end{cases}
\label{eq:matter_damping}
\end{align}

Fig. 1 shows the effects of nonzero neutrino mass on different power spectra in linear theory. On the top panel of Fig. 1, we plot the actual matter power spectrum for a cosmology with $M_\nu = 0.09\,$eV at $z=0$. On the bottom panel we plot the ratios of various power spectra in this cosmology to the matter power spectrum in a massless neutrino cosmology. To isolate the effect of the neutrino masses, we have kept fixed the value of $\Omega_{\rm m}$, $\Omega_{\rm b}$, $h$, $n_s$ and $A_s$. Note that in the massless neutrino cosmology $\delta_m = \delta_c$ by definition. As can be seen from the different curves on the bottom panel, massive neutrinos damp the total matter power spectrum with a different amplitude and shape than the CDM power spectrum.

Eq.s \ref{eq:Pmk}, \ref{eq:matter_damping}, and \ref{eq:cdm_damping} suggest that a good way to constrain neutrino masses in cosmology would be to compare the amplitude of the power spectrum at scales much larger than the free streaming scale to the one at small scales (where neutrinos perturbations have been washed out) \cite{Seljak:2006bg}. In principle, the first measurement can be obtained with Cosmic Microwave Background (CMB) data, while the latter information is present in the data from late time galaxy clustering, galaxy lensing, CMB lensing, and Lyman-alpha. Although promising, measurements of neutrino masses through relative amplitude of the power spectrum are very challenging, and they are ultimately limited by degeneracy with other cosmological parameters and unmitigated systematic effects.

As mentioned in Section 1, a more unique signature of neutrino masses is instead the scale-dependent variation of the power spectrum on $0.01<k<0.1\,\kMpc$ scales. On scales much larger than $k\sim 0.01\,\kMpc$, i.e. above the largest free streaming scale of the neutrinos over the history of the Universe, neutrinos and CDM behave similarly - i.e. they produce no change to the shape of the power spectrum. On scales smaller than $k\sim 0.1\,\kMpc$, the neutrino fluctuations are so tiny that they are essentially negligible, leading to the nearly scale-independent flattening of the power spectrum seen in the bottom panel of Fig. \ref{fig:pk}.  While changes in other cosmological parameters, e.g. a change in $\sigma_8$, can mimic the suppression of the matter power spectrum on small scales, standard $\Lambda$CDM cosmology cannot produce this scale-dependent variation in the range $0.01<k<0.1\,\kMpc$. However, in this regime, the suppression is very small and therefore extremely difficult to detect in real data. This is, of course, directly related to the value of $\fnu$ being very small for realistic neutrino masses. 

A further complication in inferring the neutrino masses from cosmology is that we cannot observe the CDM or neutrino field directly. Instead, we observe biased tracers of the underlying density fields, \eg galaxies in spectroscopic and photometric  surveys. In a cosmology with massive neutrinos and in the linear regime, the most generic bias expansion for a matter tracer $x$ is
\begin{align}
\label{eq:bx}
\delta_x = b_c \delta_c + b_\nu \delta_\nu
\end{align}
and a priori there is no reason to drop any of the two terms. Recently \citep{Villaescusa-Navarro:2013pva,Castorina:2013wga,Castorina:2015bma} have shown that due to the large hierarchy between the free streaming scale, $\lambda_{\text{fs}}(m_\nu=0.1\,\text{eV},z=0)\simeq40\,\Mpc$, and the scale $R^*\simeq1\,\Mpc$ relevant for the formation of halos and galaxies, the second term in \eq{eq:bx} is negligible compared to the first one and we can approximate
\begin{align}
\label{eq:bxc}
\delta_x \simeq b_c\delta_c\,.
\end{align}
In other words, the contribution of massive neutrinos to the mass and collapse history of halos and, consequently, galaxies is negligible. This scenario has indeed been confirmed to better than a \% level by measurement of halo bias in N-body simulations that include massive neutrinos, and linear bias is approximately scale independent if defined with respect to the CDM field \cite{Castorina:2013wga,FVN18}. This fact has interesting consequences for the abundance of massive clusters, higher point function of the density fields, and the property of cosmic voids.

Regardless of the large hierarchy between the neutrino free-streaming scale and the non-linear collapse scale, it is possible to compute corrections to \eq{eq:bxc} \cite{LoVerde:2014pxa,Chiang:2017vuk,Chiang:2018laa} that will therefore be scale-dependent. 
The analytical calculation of \cite{LoVerde:2014pxa,Munoz:2018ajr,Fidler2018}, as well as the results from N-body simulations \cite{Chiang:2018laa} indicate that $b_\nu \lesssim \mathcal{O}(1) \fnu$ and is only important on linear scales $k \le 0.01 \,\kMpc$.
Being so small, this form of scale-dependent bias is not expected to add much in terms of constraining power on massive neutrinos. It has also been shown that for voids defined using $\delta_m$ rather than $\delta_c$, there is scale dependence in the bias of such voids on large scales \cite{Banerjee:2016zaa}. However, these voids have to be identified using weak lensing measurements, where the signal to noise ratio per void is too small to have robust void definitions. Another possible approach is to use the different scale dependence in a measurement of $P_{xx}(k)=b^2\Pcc$ and of $P_m(k)$ with weak lensing, to measure neutrino masses through the ratio $P_{cc}(k)/P_{mm}(k)$. This was discussed in \citep{Schmittfull:2017ffw}, but the authors find that the constraints on the neutrino mass from the scale-dependent piece is much smaller than the total constraints coming from the damping of the power spectrum on small scales. 

The biggest impediment to the observation of the scale-dependent linear bias effect in massive neutrino cosmology, therefore, is that the signal is small. In the rest of this paper, we explore how the use of the halo environment, as defined by the total matter and CDM density fields, can enhance the size of the scale-dependent signal, and provide a new robust observable to constrain neutrino masses.

\hfill

\section{Analytical considerations on environmental bias}
\label{sec:analytical}
What physical phenomena determine the exact value of the bias parameter of a generic tracer of the LSS is an open problem in cosmology. We know that mass for halos and luminosity for galaxies are the primary driver in setting the value of the bias, but other sources of secondary bias are actively being explored.
As gravitational collapse happens locally in time and space, we expect the bias to be a function of the density of all independent fields and of (the second derivative of) the gravitational potential, along with their spatial derivatives, evaluated at the scale relevant for the process \cite{Desjacques18}. 
Since we are not interested in the bias expansion on a object-by-object basis, we also need to know the statistical properties of all the variables relevant for the collapse of halos.

The first bias model was presented more than thirty years ago by Kaiser in \citep{Kaiser84}.
In this model, halos of mass $M$ are identified as regions in the initial linear field whose density contrast field $\delta$, smoothed on the scale $R_p=(3M/4\pi)^{1/3}$, is larger than the critical value required by a simple spherical collapse model. Halo bias is then computed as the ratio between the correlation function of these constrained regions and the underlying dark matter in the limit that these halos are very rare fluctuations.
Over the years, several layers of complexity have been added to the initial model, with the inclusion of the peak constraint \cite{BBKS1986,Desjacques10}, \ie halos preferentially form at peaks in the initial field, the excursion set constraint to solve the cloud in cloud problem \cite{BCEK91,MussoSheth}, and more realistic collapse models, \eg halos are likely to form in a ellipsoidal fashion \cite{ShethTormen,Esptau}.
The predictions of this class of models, called Lagrangian bias models, are in very good agreement with bias parameters measured directly in N-body simulations that solve the exact gravitational dynamics \cite{Biagetti:2014pha,Lazeyras,Modi}.
In this picture, halos are identified as a set of constraints $\{\mathcal{C}\}$ we impose on the initial matter fields on one or more Lagrangian scales. For instance, in spherical collapse models, halos are regions of size $R_p$ in which the linearly extrapolated density contrast exceeds the critical threshold $\delta_{\rm crit}=1.686.$ In this language, halo formation is therefore identified with a selection process.
The linear bias is then defined in the following way
\begin{align}
\label{eq:bDef}
b(k) =\frac{\avg{\delta_c(k) \,|\,\{\mathcal{C}\}}}{\avg{\delta_c(k) \delta_c(k)}}\,,
\end{align}
which is the cross-correlation between the CDM and center of the constrained region divided by the CDM power spectrum. 

The linear bias discussed will computed in the initial Lagrangian Space, and therefore cannot directly be compared with measurements in simulations at low redshift, that are carried out in Eulerian Space. A mapping between Lagrangian and Eulerian coordinates exists, and it depends on the halo dynamics and the statistical properties of the nonlinear Eulerian fields. The  statement that results derived in Lagrangian space survive to Eulerian space is non-trivial. For the specific case of environmental bias we are interested in, this has been extensively checked to hold in the literature.  We refer the reader to \cite{ShiSheth,Pujol17} for more details \footnote{We have also checked that our measurements in the simulations agree with the results in \cite{Pujol17}.}. 

For Gaussian random fields as the initial matter fields, the bias can only depend on the spectral moments 
\begin{align}
\sigma^2_n(P_{ab}(k);R_i,R_j) \equiv \sigma_{n;ab;ij}^2 = \int \frac{\mathrm{d}k}{2\pi^2}\,k^{2(1+n)} P_{ab}(k)W(kR_i)W(k R_j) \, ,
\end{align}
with $a,b=\{c,\,\nu\}$, for CDM and neutrinos, $R_{i,j}$ are the scales at which we impose the constraints, and $W(kR)$ is the smoothing window that defines the constrained region. Unless otherwise noted, in the remainder of this work we will use Top-Hat window functions.

\subsection{Environmental Lagrangian bias in $\Lambda$CDM cosmologies}
\begin{figure}
\centering
\includegraphics[width=0.45\textwidth]{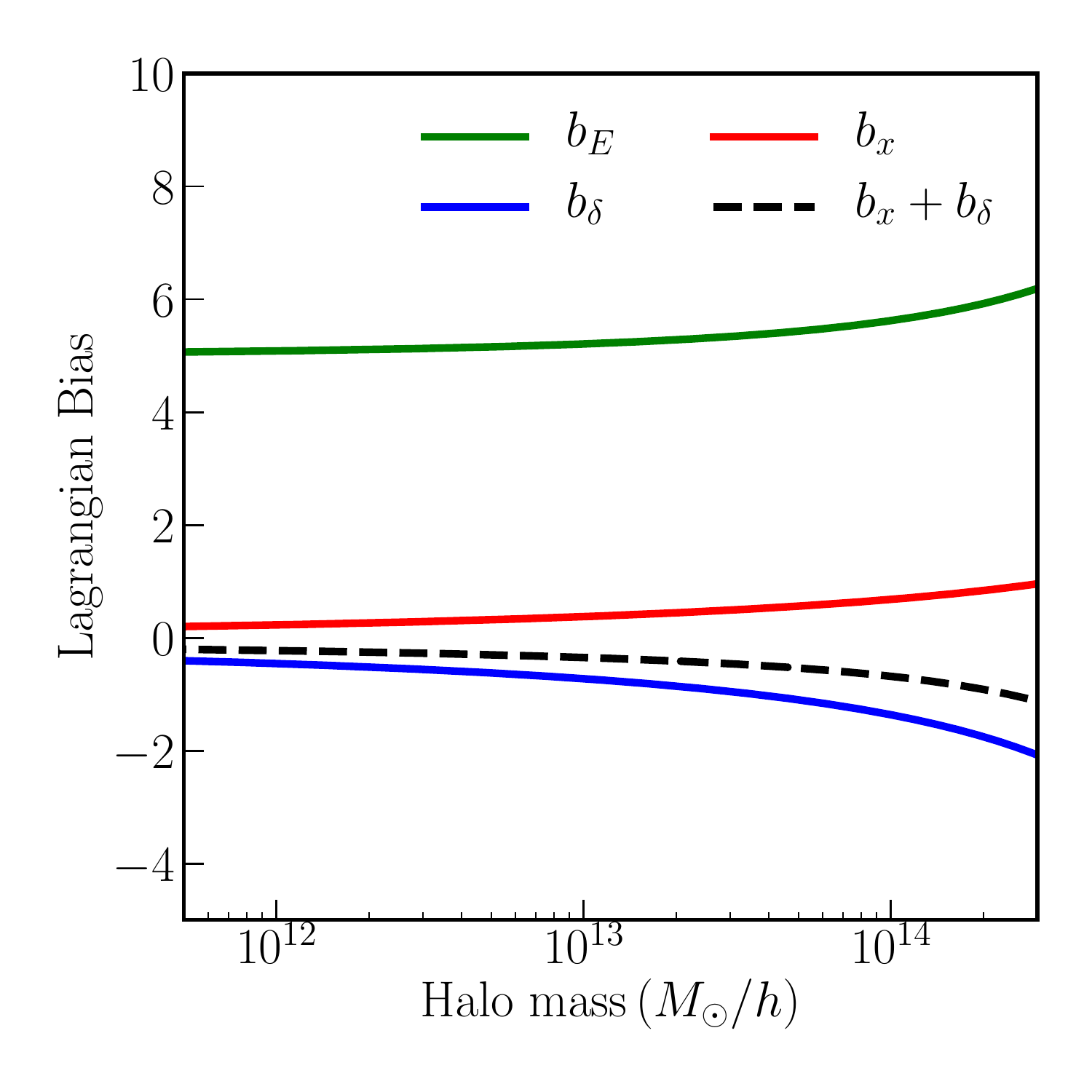}
\includegraphics[width=0.45\textwidth]{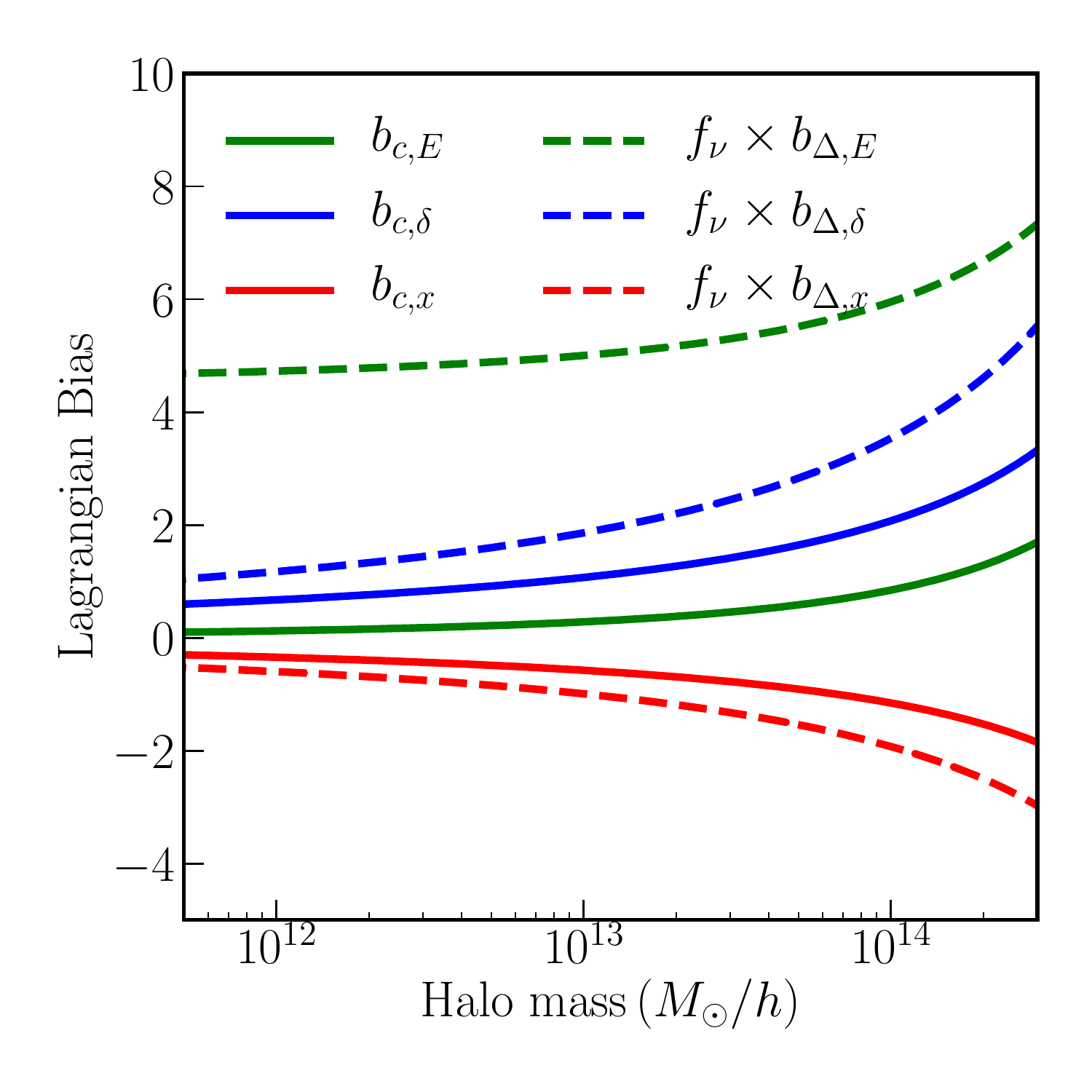}
\caption{\emph{Left Panel:} In a $\Lambda$CDM cosmology, the contributions to the Lagrangian bias of halos for a spherical environment defined at $R_E=20\,\Mpc$. The standard BBKS terms are shown in blue and red and are always subdominant compared to the environmental term in green. \emph{Right Panel:} In a cosmology with massive neutrinos of total mass $M_\nu = 0.15$ eV, the different contributions to the scale independent bias $b_c$ (continuous lines), and to the scale dependent bias $b_\Delta$ (dashed lines), produced by an environmental selection based on the value of $\Delta = \delta_m - \delta_c$ computed in a $R_E=20\,\Mpc$ sphere around each halo.}
\label{fig:b_lag}
\end{figure}

In the definition of linear bias of \eq{eq:bDef}, we did not have to specify the constraints, which can therefore incorporate any other properties of the selected sample.
For instance we can ask the question of what is the bias of halos that reside in a over/under-dense environment, the latter defined at some large scale $R_E>R_p$ \cite{AbbasSheth}. In our formalism this is expressed as 
\begin{align}
b(k) = \frac{\avg{\delta_c(k) \,|\,\{\mathcal{C}(R_p)\,,\mathcal{E}(R_E) \}}}{\avg{\delta_c(k) \delta_c(k)}}\,, 
\end{align}
in which we explicitly distinguish between the constraints $\mathcal{C}(R_p)$ we impose at the halo scale $R_p$, and the ones on the environment, $\mathcal{E}(R_E)$, at scale $R_E$. In this section we compute the bias of halos as a function of environment. For simplicity, we restrict our analysis to a $\Lambda$CDM cosmology. Our calculation generalizes the results of \citep{AbbasSheth,CastorinaSheth,Pujol17,ShiSheth} to more realistic halo definitions. We assume that $\mathcal{C}$ is a function of the density contrast of CDM smoothed at halo scale, $\delta_c(R_p) \equiv \delta_{c,p}$, its second spatial derivative, \ie the peak curvature $\nabla^2\delta_{c,p}$, and that $\mathcal{E}$ depends on the value of the CDM overdensity constrast at the environmental scale $R_E$, $\delta_{c,E}$. For shortness of the presentation we will not include dependence on the tidal fields in the equations below, but it is included in the numerical evaluations following \cite{Castorina16,Esptau}. 
Since CDM is the only relevant field in $\Lambda$CDM cosmology, we will drop the subscript $c$ in the remainder of this section.
We thus need to keep track of the spatial correlation of three degrees of freedom, which can be arranged in a 3-dimensional Gaussian distribution.
It is convenient to define standardized variables with unit variance \cite{BBKS1986,AbbasSheth}
\begin{align}
\nu_{p} \equiv \frac{\delta_p}{\sigma_{0;pp}}\quad,\quad x_{p} \equiv -\frac{\nabla^2\delta_p}{\sigma_{2;pp}}\quad,\quad \nu_{E} \equiv \frac{ \delta_E}{\sigma_{0;EE}}\;,
\end{align}
that jointly form a 3-dimensional Gaussian distribution with zero mean and covariance
\begin{align}
\Sigma \equiv \left(
\begin{array}{ccc}
 1 & \gamma _x & \gamma _E \\
 \gamma _x & 1 & \epsilon\\
 \gamma _E & \epsilon & 1 \\
\end{array}
\right) \, ,
\end{align}
with the following definitions
\begin{align}
&\gamma_x \equiv \frac{\sigma_{1;pp}^2}{\sigma_{0;pp} \,\sigma_{2;pp}}\quad,\quad  \gamma_E\equiv \frac{\sigma_{0;pE}^2}{\sigma_{0;pp}\, \sigma_{0;EE}}\quad\,\epsilon\equiv \frac{\gamma_x}{\sigma_{0;EE}}\frac{\mathrm d \sigma_{0;pE}^2}{\mathrm d \sigma_{0;pp}} \, .
\end{align}
The off-diagonal entries of $\Sigma$ describe the cross correlation between the density field at the halo/peak scale and the curvature field $x_p$, $\gamma_x$, the correlation between the former and the environment, $\epsilon$, and the correlation between the density field at the environmental and halo scale, $\gamma_E$.
Within this model, by using simple properties of Gaussian random fields and linear algebra one arrives to the following expression for the Lagrangian linear bias of environmentally selected dark matter halos
\begin{align}
\label{eq:bcE}
b &= \frac{\avg{\delta | \mathcal{C}(\nu_p,x_p;R_p)\,,\mathcal{E}(\nu_E;R_E)}}{\avg{\delta \delta}} \\
& =\frac{\avg{\delta \nu_p}}{\avg{\delta \delta}} [(1-\epsilon^2)\avg{\nu_p| \mathcal{C}\,,\mathcal{E}} +\left( \gamma_x \epsilon-\gamma_E\right)\avg{\nu_E| \mathcal{C}\,,\mathcal{E}} + (\epsilon_c \gamma_E-\gamma_x)\avg{x_p| \mathcal{C}\,,\mathcal{E}}]/A \notag\\
&+\frac{\avg{\delta \nu_E}}{\avg{\delta \delta}} [(\gamma_x \epsilon-\gamma_E)\avg{\nu_p| \mathcal{C}\,,\mathcal{E}}+(1-\gamma_x^2)\avg{\nu_E| \mathcal{C}\,,\mathcal{E}}+(\gamma_x\gamma_E-\epsilon)\avg{x_p| \mathcal{C}\,,\mathcal{E}}]/A \notag \\
&+\frac{\avg{\delta x_p}}{\avg{\delta \delta}} [(\gamma_E \epsilon-\gamma_x)\avg{\nu_p| \mathcal{C}\,,\mathcal{E}}+(\gamma_x \gamma_E-\epsilon)\avg{\nu_E| \mathcal{C}\,,\mathcal{E}}+(1-\gamma_E^2)\avg{x_p| \mathcal{C}\,,\mathcal{E}}]/A \label{eq:bcxE} \, ,
\end{align}
with
\begin{equation}
A =|\Sigma|= 1-\gamma_x^2-\gamma_E^2-\epsilon^2+2 \gamma_x \gamma_E \epsilon\,.
\end{equation}
Notice that since $x_p$ is proportional to the second derivative of the density field, the bias in the above expression will be $k
$-dependent \cite{BBKS1986,MussoSheth,Desjacques10}. In the case of no environmental constraint, $\epsilon=\gamma_E=0$, we recover the well-known results of BBKS \cite{BBKS1986} 
\begin{align}
\avg{\delta|\mathcal{C}} = \avg{\delta \delta_p}\frac{\avg{\nu_p-\gamma_x x_p|\mathcal{C}}}{\sigma_p(1-\gamma_x^2)} + \avg{\delta x_p}\frac{\avg{x_p-\gamma_x \nu_p|\mathcal{C}}}{1-\gamma_x^2} \, .
\end{align}
On the other hand, if we neglect the peak constraint on the second derivative of the density field, \ie $\mathcal{C}=\mathcal{C}(\nu_p,R_p)$ only and $\gamma = 0$, we obtain the expressions in \cite{ShiSheth}
\begin{equation}
\avg{\delta|\mathcal{C}\,,\mathcal{E}} = \avg{\delta \delta_p}\frac{\avg{\nu_p-\gamma_E \nu_E|\mathcal{C}\,,\mathcal{E}}}{\sigma_p(1-\gamma_E^2)} + \avg{\delta \delta_E}\frac{\avg{\nu_E - \gamma_E\nu_p|\mathcal{C}\,,\mathcal{E}}}{\sigma_E(1-\gamma_E^2)}\,.
\end{equation}
This result is a consequence of the statistical properties of Gaussian fields, and shows that the bias parameters are only functions of the constrained variables \cite{Castorina16}.
In the low-$k$ limit in which $k\ll R_p^{-1},\,R_E^{-1}$, the contribution to the bias of the third line of \eq{eq:bcxE} can be dropped and $\avg{\delta \delta_p} = \avg{\delta \delta_E} =\avg{\delta \delta}$, such that the scale independent piece of linear bias can be rearranged as 
\begin{align}
\label{ew:bE}
b = b_\delta \avg{\nu_p| \mathcal{C}\,,\mathcal{E}} + b_x \avg{x_p| \mathcal{C}\,,\mathcal{E}}+b_E \avg{\nu_E| \mathcal{C}\,,\mathcal{E}}\,,
\end{align}
where we have combined all terms that multiply the same conditional expectation values in \eq{eq:bcxE}. The definitions of $b_\delta,b_x\,b_E$ can therefore, be simply read off by comparing \eq{ew:bE} to \eq{eq:bcxE}. In our model, the mean value of the density $\nu_p$, and of the curvature, $x_p$, at the halo scale and the mean environmental density, $\nu_E$  determine the bias. The values of $b_\delta$ (blue), $b_x$ (green) and $b_E$ (red) as a function of halo mass, $M=4\pi/3 \bar{\rho}R_p^3$, are shown in Fig. \ref{fig:b_lag} for an environmental scale of $R_E=20\,\Mpc$ at $z=0$.
We find that once we impose the additional environmental constraint, then $b_E> b_\delta\,,b_x$ by factors of $5$ or more. Also, the two standard terms $b_\delta$ and $b_x$ are not very sensitive to halo mass once we include an environmental selection.
On scales where $R_p\simeq R_E$, the density field at $R_E$ cannot be no longer be interpreted as the halo environment and our calculation breaks down.

Since both $\avg{\nu_p| \mathcal{C}\,,\mathcal{E}}$ and $\avg{x_p| \mathcal{C}\,,\mathcal{E}}$ are $\mathcal{O}(1)$ numbers \citep{BBKS1986,Esp,Esptau} and $b_\delta$ and $b_x$ have opposite signs, we find that the bias is, to a very good approximation, determined by the value of the environmental density around halos, $\avg{\nu_E| \mathcal{C}\,,\mathcal{E}}$, \textit{almost independently of halo mass}
\begin{align}
b\simeq b_E \avg{\nu_E| \mathcal{C}\,,\mathcal{E}}\,.
\end{align}
This result is a consequence of the shape of the initial linear power spectrum for which $\sigma_E\gg\sigma_p$ if $R_p<R_E$.
Our findings are in agreement with \citep{AbbasSheth,Pujol17,ShiSheth,ZeroBias}, and qualitatively explain a series of results concerning assembly bias of dark matter halo in different environment \citep{Salcedo2018,Han2018,Ramakrishnan}. 

At this point it is worth reminding again that our calculation of linear bias has been carried out in the initial Lagrangian field, but the environment is usually defined in the late time Eulerian field where observations are made. While this introduces quantitative differences in the calculation, the picture described above still holds \cite{Pujol17,ShiSheth}, and we will use it to understand the clustering pattern in cosmologies with massive neutrinos.

\subsection{Scale-dependent environmental bias}
In a Universe filled with massive neutrinos, along with CDM, we have several ways of  defining the environment. 
In the analysis of the previous section, the results would not change if we replace  the CDM field in the definition of the environment with $\delta_\nu$ or $\delta_c$, or any linear combination of the two. This means that, despite the fact that we can assume to a very good approximation mass selected halos follow \eq{eq:bxc}, adding an environmental constraint that depends on the density field of neutrinos, the halo bias can be made scale-dependent with respect to CDM.
Since we have shown that the environment is the strongest effect in setting the amplitude of the bias, in the trivial case the environment is defined with respect to the neutrino field, we expect to see large scale dependence and all the results from the previous section carry through.
In real data, the neutrino field is not directly observable and we will have to use a combination of the CDM field measured by dark matter halos and the total matter field measured for instance with weak lensing data.
In this section we will explore a few possibilities to define the halo environment that introduce  scale-dependent bias on linear scales.

\subsubsection{Total matter environment}
It is straightforward to generalize the calculation of the Lagrangian bias of halos described in the previous section to the case of the environment defined in terms of the total matter field, \ie the mass weighted sum of CDM and neutrino density fields.
In the low-$k$ limit, $k\ll R_p^{-1}$ and $k\ll R_E^{-1}$, the linear bias can now be written
\begin{align}
\label{eq:bmE}
b(k) \equiv \avg{\delta_c | \mathcal{C}\,,\mathcal{E}(\delta_m)}/ \avg{\delta_c \delta_c}= b_c  +  b_{\nu} P_{c\nu}(k)/P_{cc}(k) \,, 
\end{align}  
with $\mathcal{C}$ the halo constraints depending only on CDM, and $\mathcal{E}$ the environmental constraints which depends on $\delta_m = f_c \delta_c + f_\nu \delta_\nu$. By repeating the steps in the previous section it is easy to see that the amplitude of the scale-dependent piece of the bias is proportional to the difference of the mean environmental total matter and cold dark matter density
\begin{align}
b_{\nu} \simeq \avg{\nu _{m,E} -\nu_{c,E}|\mathcal{C},\mathcal{E}} b_E \,,
\end{align}
where as before the mass dependence of the $b_E$ term is very weak.
To $\mathcal{O}(f_\nu)$ we find
\begin{align}
\nu_{m,E} =\frac{\delta_{m,E}}{\sigma_{0;mm;EE}} \simeq \frac{\delta_{c,E}}{\sigma_{0;mm;EE}}-f_\nu\frac{(\sigma_{0;cc;EE}) ^2\delta_{\nu,E}-(\sigma_{0;c\nu;EE} )^2\delta_{c,E}}{(\sigma_{0;cc;EE}) ^3} + \mathcal{O}(f_\nu^2) \,,
\end{align}
such that 
\begin{align}
\avg{\nu _{m,E} -\nu_{c,E}| \mathcal{C},\mathcal{E}} \simeq f_\nu\frac{(\sigma_{0;cc;EE}) ^2\avg{\delta_{\nu,E}|\mathcal{C},\mathcal{E}}-(\sigma_{0;c\nu;EE} )^2\avg{\delta_{c,E}|\mathcal{C},\mathcal{E}}}{(\sigma_{0;cc;EE})^3} \,.
\end{align}
On most environmental scales, $R_E$, the first term is bigger than the second one, which means we are effectively constraining the neutrino environment. On the other hand, we have an effect even if the regions we select have $\avg{\delta_{\nu,E}|\mathcal{C},\mathcal{E}}\simeq 0$. This term arises since selecting only halos residing in regions with zero environmental neutrino overdensity is still a constraint that will produce a $k$-dependent contribution to the bias.
Unfortunately, the scale dependence of the bias introduced by the environmental constraint on the total matter density is proportional to $f_\nu$ and hence very small. This result is aniticipated, becuase without a measurement of the environmental CDM density, one cannot distinguish CDM and total matter better than $\mathcal{O}(f_\nu)$.

\subsubsection{Environmental constraint on the CDM and total matter}
The analysis of the previous section on the bias of halos selected based on their  total matter environment suggests that in order to get rid of the small parameter $f_\nu$ and probe the neutrino environment directly, we need to independently measure both the CDM and total matter environment. Obviously in the N-body simulations one can directly measure the neutrino field, but this is not possible in data from cosmological surveys. While the total matter field can be measured from weak lensing measurements, one needs to infer the CDM field from the clustering of galaxies - where uncertainty in bias and the presence of stochasticity can play a major role.

It is easy to write the more general case of joint CDM and total matter environmental constraints, but for brevity of the presentation we focus on the case the environment depends on difference between the two  fields
\begin{equation}
\Delta (R_E)\equiv \delta_{m,E} - \delta_{c,E} = f_\nu (\delta_{\nu,E}-\delta_{c,E})\,,
\end{equation}
as it will allow us to check the  non trivial scale-dependent effects predicted by our model. The expression in \eq{eq:bmE} already contains the most general scale dependence allowed in the two species scenario, but for our purposes we rewrite it as 
\begin{align}
\label{eq:bD}
b(k) = \avg{\delta_c | \mathcal{C,E}(\Delta)}/\avg{\delta_c \delta_c} =& b_c + b_\Delta P_{c\Delta}(k)/P_{cc}(k) \\ \nonumber  =& b_c + b_\Delta f_\nu (P_{c\nu}(k)/P_{cc}(k)-1) \,.
\end{align}
We can decompose both $b_c$ and $b_\Delta$ in the same way as in \eq{ew:bE}, in which the value of the bias parameter depends on the average of three relevant variables, $\nu_p$, $x_p$ and $\Delta$. In the right panel of Fig. \ref{fig:b_lag} we plot the three contributions to $b_c$ (solid lines), and the three contributions to $b_\Delta$ multiplied by $f_\nu$ (dashed lines),  for $R_E = 20\,\Mpc$ and a total value of neutrino masses of $M_\nu = 0.15$ eV. 
Similar to the $\Lambda$CDM scenario, the environmental constraint is the strongest, such that $f_\nu b_\Delta > b_c$ and  the scale dependent piece of the bias dominates over the scale independent one. The value of $b_\Delta$ depends on how the environment correlates with the local halo density and curvature, the dashed blue and red lines, but to a very good approximation is just proportional to the mean value of the environmental density $\avg{\Delta| \mathcal{C},\mathcal{E}}$, shown as the green dashed line.
\eq{eq:bD} implies that on scales much larger than the free streaming scale where $P_{c\nu}\simeq P_{cc}$ it is virtually impossible to separate the effect of the environment from the standard scale independent bias $b_c$.
We will see in the next section that this feature is confirmed in N-body simulations, and it is quantitatively different from effects of an environment defined w.r.t. the neutrino field or any other combination of dark matter and total matter fields, although formally they can all be written using \eq{eq:bmE}.

Assuming the value of $b_\Delta$ is primarily set by the environmental constraint we then arrive to 
\begin{align}
b(k) &= b_c +f_\nu b_\Delta (P_{c\nu}(k)/P_{cc}(k)-1) \\
     & \simeq b_c + f_\nu b_{\Delta,E} \frac{\avg{\Delta\,|\,\mathcal{C,E}}}{\sigma_{0;\Delta\Delta}^2} (P_{c\nu}(k)/P_{cc}(k)-1)  \notag \\
     & =b_c+b_{\Delta,E}\frac{\avg{\delta_\nu-\delta_c \,|\,\mathcal{C,E}}}{\sigma_{0;cc;EE}^2+\sigma_{0;\nu\nu;EE}^2-2\sigma_{0;c\nu;EE}^2} (P_{c\nu}(k)/P_{cc}(k)-1) 
\end{align}
where we notice that all powers of $f_\nu$ have dropped from the final expression. This result confirms the intuition of the previous section, and we have therefore identified infinite possible environment definitions which all lead to very large scale dependence proportional to the neutrino linear transfer function. If the environment is defined at $R_E\ll \lambda_{\rm fs}$ then we can further simplify the expression for the bias and arrive to
\begin{align}
b(k) \simeq b_c-b_{\Delta,E}\frac{\avg{\delta_{c,E} \,|\,\mathcal{C,E}}}{\sigma_{0;cc;EE}^2} (P_{c\nu}(k)/P_{cc}(k)-1) \,.
\label{eq:sd_bias}
\end{align}
Therefore, these analytical calculations show that the bias of halos selected using certain combinations of the CDM and matter environment should have a scale independent term, i.e. the first term on the RHS of Eq. \ref{eq:sd_bias}, and a scale-dependent term, i.e. the second term on the RHS of Eq. \ref{eq:sd_bias}. Importantly, apart from a normalization factor, the dependence of the second term on scale is set only by the ratio of the transfer functions of the neutrinos and CDM - a quantity which depends directly on the neutrino mass. 
Within the toy model presented in this section both the value of $b_{\Delta,E}$ and of $\avg{\delta_{c,E} \,|\,\mathcal{C,E}}$ can be predicted analytically. In order to do the same in real data one would need a map from Lagrangian to Eulerian coordinates at the fields level, as well as the full non-linear probability distribution function (pdf) of the Eulerian environmental variables. In a N-body simulation this can be done, see for instance \cite{Pujol17} for a discussion of environmental bias in $\Lambda$CDM cosmologies, but for observed galaxy samples this could be quite challenging. Since we are solely interested in the scale dependence of the bias as a possible way of constraining neutrino masses, in the next sections we will therefore just fit for a free parameter for the amplitude of the second term in \eqs{eq:bmE}{eq:sd_bias}.

\section{Environmental scale-dependent bias in simulations}

\begin{figure}
\centering
\includegraphics[width=0.5\textwidth]{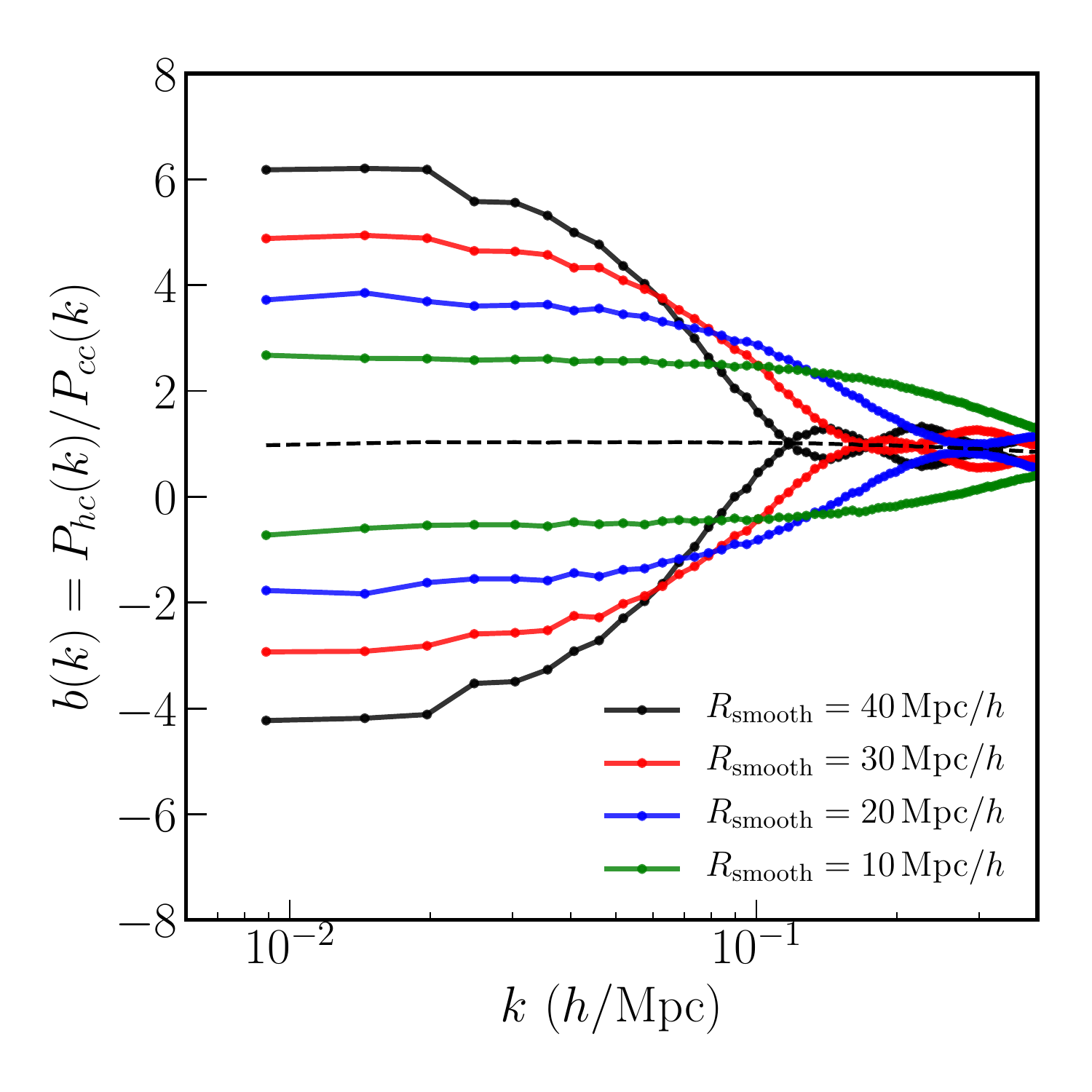}
\caption{We consider all halos with masses above $4\times 10^{11}M_\odot/h$ for the cosmology with $M_\nu = 0.10\,$eV neutrinos at $z=0$. Their halo bias is shown with the black dashed line. We then split the halos into two populations according to the CDM overdensity at a scale $R_{\rm smooth}$: 1) halos where the CDM overdensity is above the median (upper lines), and 2) halos where the CDM overdensity is below the median (bottom lines). In all cases, for sufficiently large scales, the bias becomes scale-independent.}
\label{fig:cdm_split}
\end{figure}

We use N-body simulations of boxes with side $1\,$Gpc/h, and with $1600^3$ CDM particles and $1600^3$ neutrino particles belonging to the \textsc{HADES} simulation suite\footnote{\url{https://franciscovillaescusa.github.io/hades.html}}, the precursor of the \textsc{Quijote} simulations\footnote{\url{https://github.com/franciscovillaescusa/Quijote-simulations}} \citep{Quijote}. The simulations are run using a version of the \textsc{Gadget} code \cite{Springel:2005mi}, which has been suitably modified to include massive neutrinos \cite{Viel:2010bn}. The values of the cosmological parameters in our fiducial cosmology are: $h=0.6711$, $\Omega_b = 0.049$, $\Omega_\Lambda = 0.6825$, $A_s = 2.13\times 10^{-9}$, and $n_s = 0.9624$. We run simulations with three different neutrino masses: $M_\nu=0.08\,$eV, $M_\nu = 0.10\,$eV and $M_\nu=0.12\,$eV. The initial conditions were generated at $z=99$ using the \textsc{reps} code \cite{2016ascl.soft12022Z} to ensure that the growth factors for individual species reproduce the correct amplitude of the power spectrum at $z\sim 0$. We assume three degenerate neutrino mass species. We identify halos in the simulations using a FoF halo finder, run on the CDM particles only. Since we do not use information about the halo profiles, but just their positions and masses, we use an aggressive lower bound on the halo masses. This is enforced by tagging all objects with more than 20 simulation particles, and corresponds to a physical mass cutoff of $\sim 4\times 10^{11}M_\odot/h$.

From the simulations, we use the CDM and neutrino particle positions to create a grid of densities for each species using Cloud-in-Cell (CIC) deposition scheme. From the computed CDM and neutrino ovedensities, we also calculate the total matter overdensities. Using the halo positions in the simulation volume, we again use a CIC deposition scheme to define the halo overdensity field on the same grid. The density fields are then smoothed on some scale, $R_{\rm smooth}$, using a top-hat smoothing kernel in 3 dimensions. We use 4 different values of $R_{\rm smooth}$ to demonstrate the dependence of the effect with the scale at which we smooth the different fields. The smallest value of $R_{\rm smooth}$ considered here is $10\,$Mpc$/h$, a scale on which the smoothed fields are not expected to be completely Gaussian. The largest smoothing scale we use is $40\,$Mpc$/h$. We emphasize that all bias measurements presented in this Section are in Eulerian space, rather than in Lagrangian space as presented in Section \ref{sec:analytical}. We show below that the qualitative behavior of the measured Eulerian bias at linear scales, when conditioned on various halo environments, follows those computed in Section \ref{sec:analytical}, even though the exact mapping of values of each of the bias terms from Lagrangian space to Eulerian space may not be known.

\subsection{Using the CDM environment}
First, we study the behavior of the large scale bias when halos are selected according to their environmental properties, defined as the value of the CDM overdensity on the smoothed $R_{\rm smooth}$ scale.
To implement this, we interpolate the value of smoothed CDM overdensity field, $\delta_c$, from the grid to the position of every halo. We then find the median value of $\delta_c$, and divide the halo catalog into two - one population contains all halos which have $\delta_c \geq \delta_{c, {\rm median}}$ and the other contains all halos for which $\delta_c < \delta_{c, {\rm median}}$. Once the two populations have been defined, we compute the halo overdensity field (using a CIC deposition) of each. Next, we compute the auto power spectrum of each of the overdensity fields, as well as the cross correlation of the halo overdensity field with the CDM overdensity field. This allows us to compute the bias of the two populations, either from the cross correlation:
\begin{equation}
b(k) = \frac{P_{hc}(k)}{P_{cc}(k)}\, ,
\end{equation}
or from the autocorrelation:
\begin{equation}
b(k) = \sqrt[]{\frac{P_{hh}(k)}{P_{cc}(k)}} \,.
\end{equation}
We note that, in general, the above two definition will yield different results, even on large-scales. We show the halo bias results, computed using the cross-correlation estimator, from the $M_\nu=0.10\,$eV simulation in Fig. \ref{fig:cdm_split}.
The black dashed line indicates the bias of the full halo sample. The solid lines represent the bias of the the two populations identified by considering the CDM environment at different smoothing lengths (represented by different colored curves). As the smoothing length increases, the difference in the amplitude of the bias of the two populations also increases. However, at the largest scales accessible from our simulation volume, the bias is scale-independent for all smoothing scales, in agreement with the theory expectations from Section \ref{sec:analytical}.

As discussed in Section \ref{sec:scale_dep}, the effect of neutrinos on halo bias is expected to be slightly larger when considering the total matter field environment, $\sim \mathcal{O}(f_\nu)$. Since $f_\nu \sim 1\%$, the change in the scale-dependence of the linear halo bias is not expected to be significant. Indeed, from the measurements, we find similar results to those shown in Fig. \ref{fig:cdm_split} when we split the halos into two populations based on the value of $\delta_m$, rather than on $\delta_c$ - i.e. - the bias of the populations remains scale independent on large scales, while the actual value of the bias for each population changes from the value of the bias for the full sample.

\begin{figure}
\centering
\includegraphics[width=0.45\textwidth]{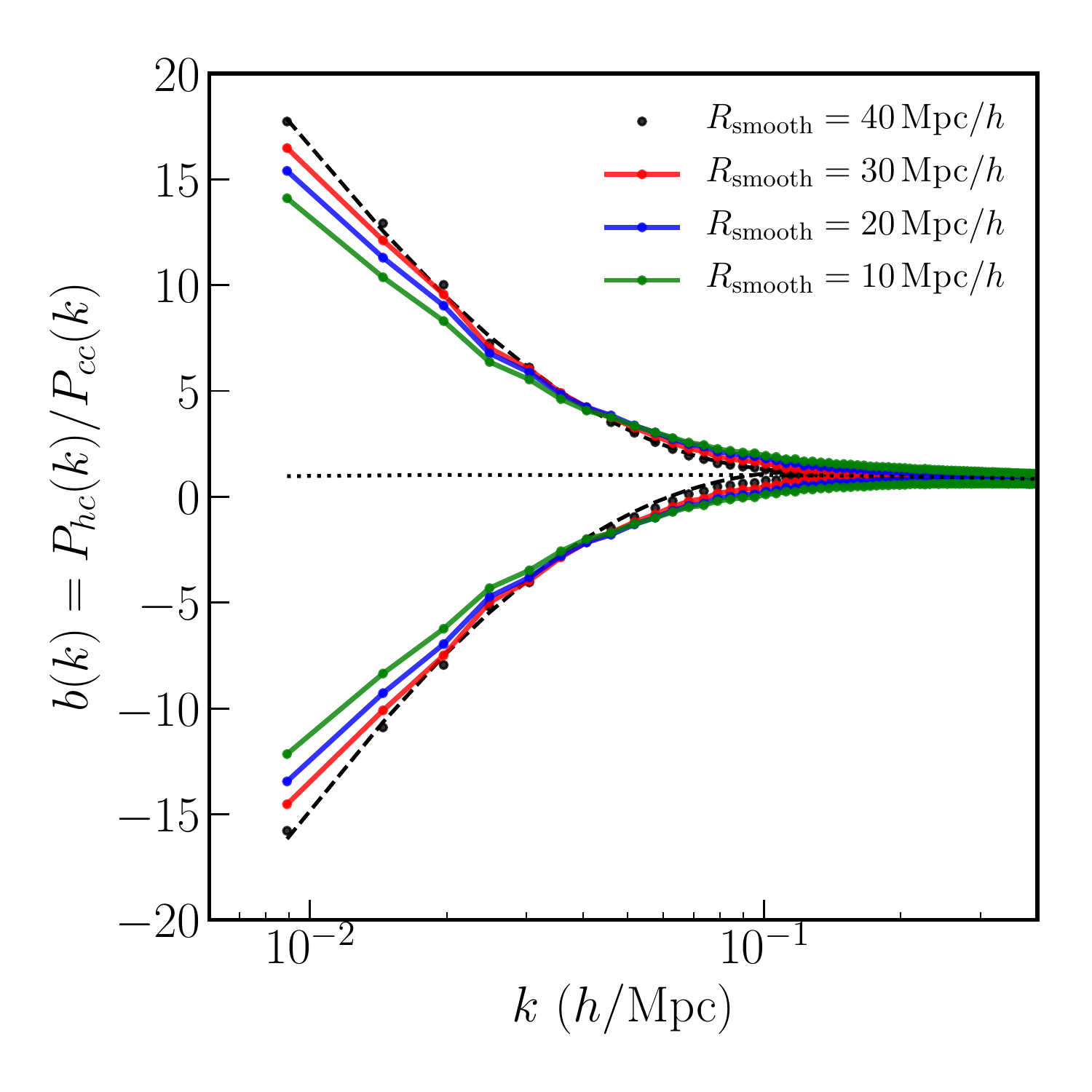}
\includegraphics[width=0.45\textwidth]{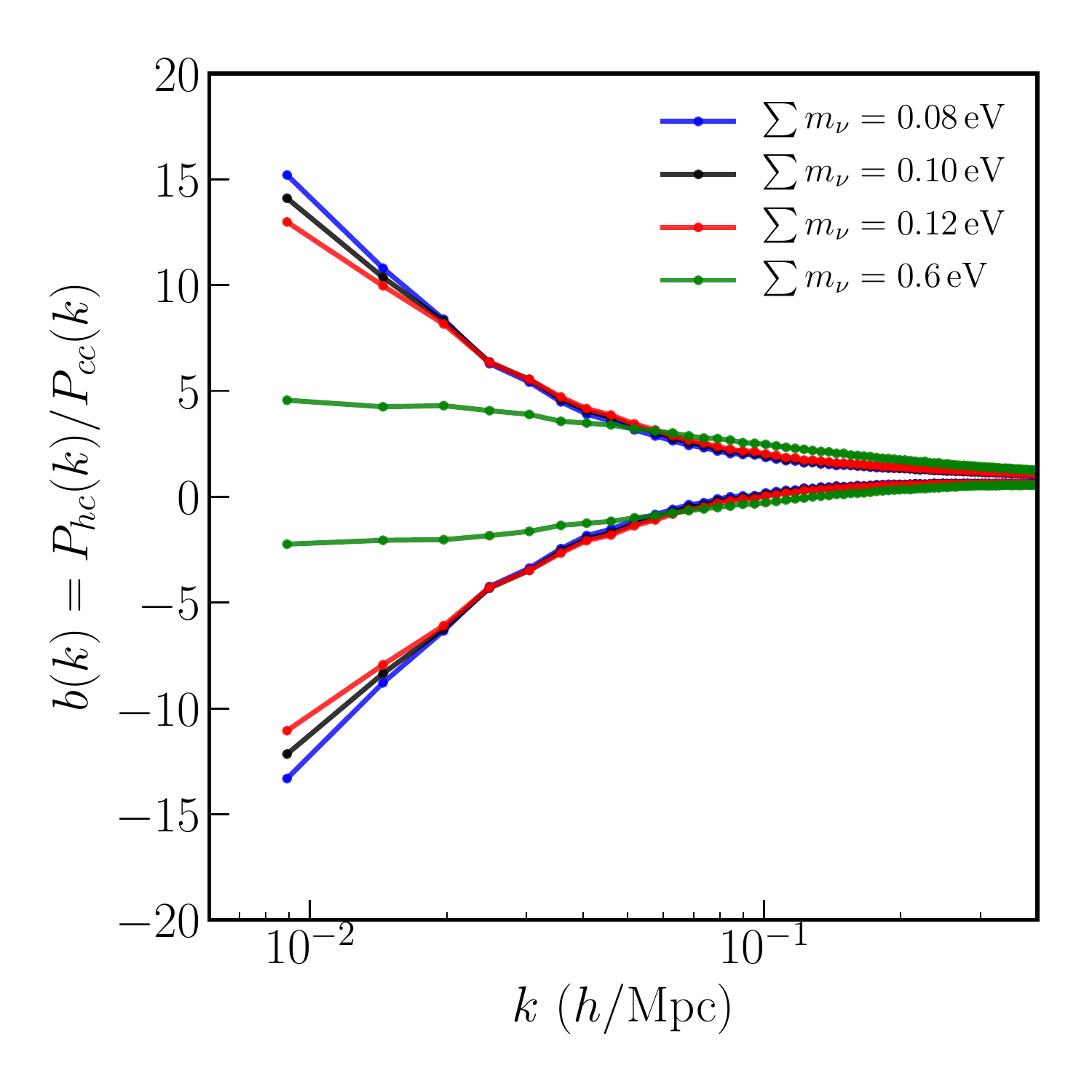}
\caption{\emph{Left panel:} Bias of the two halo populations when split on the overdensity in the neutrino field at scale $R_{\rm smooth}$ for $M_\nu = 0.10\,$eV, at $z=0$. The dotted black line indicates the bias of the full halo sample. For the smoothing scale of $40\Mpc$, we plot the fit to the data points using Eq. \ref{eq:fit_equation} with the dashed black lines. \emph{Right panel:} Bias of the two halo populations when split on the overdensity in the neutrino field at scale $R_{\rm smooth}=10\,$Mpc/h for different sum of neutrino masses. Note that when the mass of the neutrino is large, $\left(M_\nu = 0.6\,{\rm eV}\right)$, the neutrino transfer function tends towards the CDM transfer function on large scale, and the bias almost flattens to a constant value.}
\label{fig:nu_split}
\end{figure}

\subsection{Using the neutrino environment}
It is expected that the strongest signatures of neutrinos on the large scale bias will appear when we use information about the neutrino environment around each halo to perform the split into two different populations.
While the neutrino overdensity is not really an observable, we can measure it directly in the simulations. Therefore, we repeat the above procedure of splitting the halos, but now according to the value of $\delta_\nu$ smoothed on scale $R_{\rm smooth}$. The results for the bias are shown in Fig. \ref{fig:nu_split}. We now see an extremely strong scale-dependent bias on large scales that gets stronger as we increase $R_{\rm smooth}$. The behavior of the bias with $k$ is very well fit by 
\begin{equation}
    b(k) = \left(b_{c} + b_\nu \left(\frac{P_{c \nu}(k)}{P_{cc}(k)}\right)\right)W\left(kR_{\rm smooth}\right) + C \,,
\label{eq:fit_equation}
\end{equation}
where $b_{c}$, $b_\nu$, and $C$ are scale independent free parameters. $W(kR_{\rm smooth})$ is the spherical top-hat window function on scale $R_{\rm smooth}$:
\begin{equation}
    W\left(kR\right) = 3 \frac{\left(\sin\left(kR\right) + kR\cos\left(kR\right) \right) }{\left(kR\right)^3} \,.
\end{equation}
The form of the fitting function is inspired by the RHS of Eq. \ref{eq:sd_bias} and the discussion in Section \ref{sec:analytical}. Although Eq. \ref{eq:sd_bias} refers to Lagrangian space biases, we expect the functional form to also work for Eulerian space, even if the parameter values are different. Note that in Eq. \ref{eq:fit_equation}, apart from the known scale dependence coming from the window function, the shape is set entirely by the ratio of the transfer function of the neutrinos and CDM. In Fig. \ref{fig:nu_split} We plot the results of the best fit using this functional form for $R_{\rm smooth}=40\,$Mpc$/h$ (black dashed line), and similar results hold for other smoothing scales.

Further, as discussed in Section \ref{sec:analytical} and evident from \eq{eq:fit_equation}, the shape of the scale-dependent bias on large scales should change as the neutrino mass is changed. To show this, we consider simulations run with different neutrino masses, specifically $M_\nu=0.08\,$eV, $M_\nu=0.10\,$eV and $M_\nu=0.12\,$eV. We also run a simulation with extremely high neutrino mass, $M_\nu=0.6\,$eV. We fix $R_{\rm smooth} = 10\,$Mpc/h across the simulations, and split the halos from each simulation into two sets based on $\delta_\nu$. 

The large scale bias of each set are shown on the right panel of Fig. \ref{fig:nu_split}. It is interesting to note that the largest departure from scale independence is seen in the simulation with the lowest $M_\nu$. In the context of the damping of the power spectrum on small scales, the strength of the damping effect scales directly with $f_\nu$, or equivalently, with $M_\nu$. However, for the scale-dependent bias effect shown here, the size of the effect does not scale with $f_\nu$, but with ratio of the CDM and neutrino transfer functions, as discussed in Section \ref{sec:analytical}. Lowering the total neutrino mass makes the transfer function of neutrinos more different from the CDM transfer function, leading the strongest effect. For the largest neutrino masses, the bias on linear scales is almost flat, since the CDM and neutrinos transfer functions are more similar.

It is worth noting that since the shape of $b(k)$ for these populations of halos on large scales is set by the relative shapes of the neutrino and CDM transfer functions, this effect is not degenerate with the shape of the large scale bias expected from cosmologies with primordial non-gaussianties \cite{Dalal:2007cu,ZeroBias}. In the latter, the shape of the bias generally continues to scale as $k^{-2}$ up to extremely large scales. On the other hand, on scales larger than the maximum free streaming length of the neutrinos, the transfer function of the neutrinos will be identical to the transfer function of neutrinos, returning the bias to a scale independent quantity. For realistic neutrino masses, this scale is roughly a few Gpc$/h$, i.e. beyond the scales included in our simulations. However, for the $M_\nu=0.6\,$eV, this scale is $\sim 1\,$Gpc$/h$, and the right panel of Fig. \ref{fig:nu_split} does indeed show the bias becoming almost scale independent on the largest scales accessible from the simulation.

\begin{figure}
\centering
\includegraphics[width=0.4\textwidth]{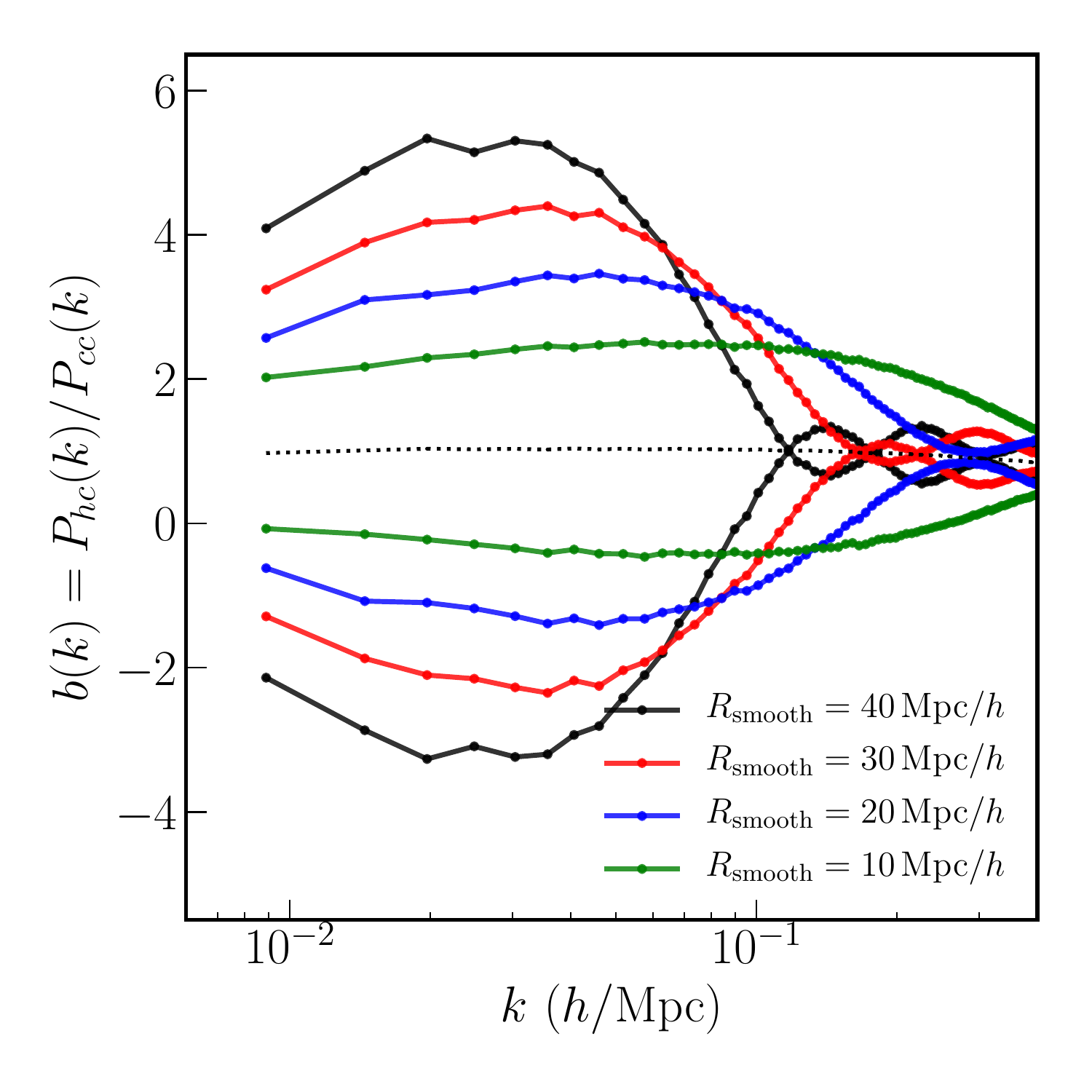}
\includegraphics[width=0.4\textwidth]{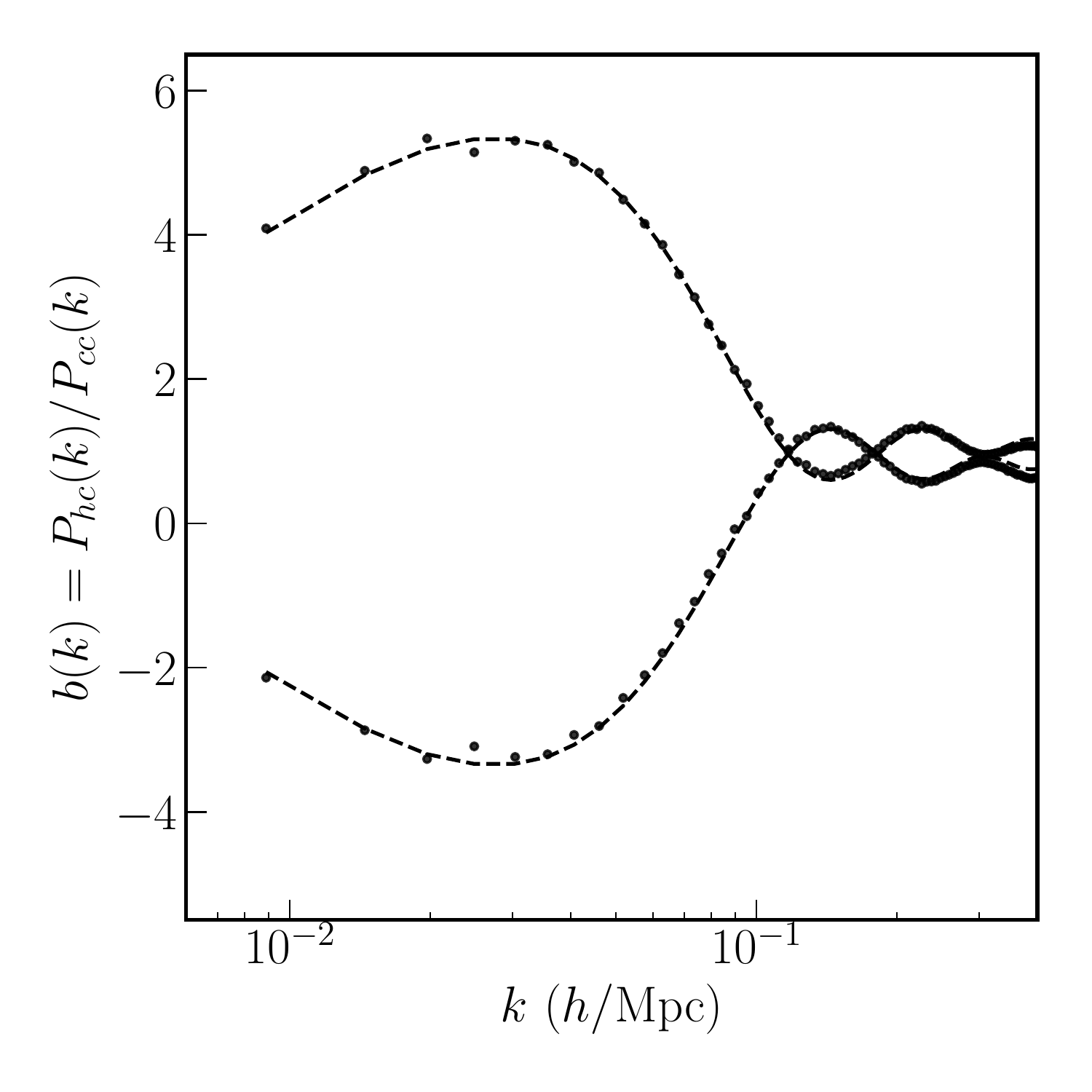}
\caption{\emph{Left panel:} Bias of the two halo populations when split on environment defined by $\delta_c - \delta_{\rm m}$ at scale $R_{\rm smooth}$ for $M_\nu = 0.10\,$eV at $z=0$. The dotted black line indicates the bias of the full halo sample.  \emph{Right panel:} For smoothing scale of $40\Mpc$, we plot the best fit curve of the form shown in Eq. \ref{eq:fit_equation} to the data using the black dashed lines.}
\label{fig:cdm_minus_matter}
\end{figure}

\begin{figure}
\centering
\includegraphics[width=0.4\textwidth]{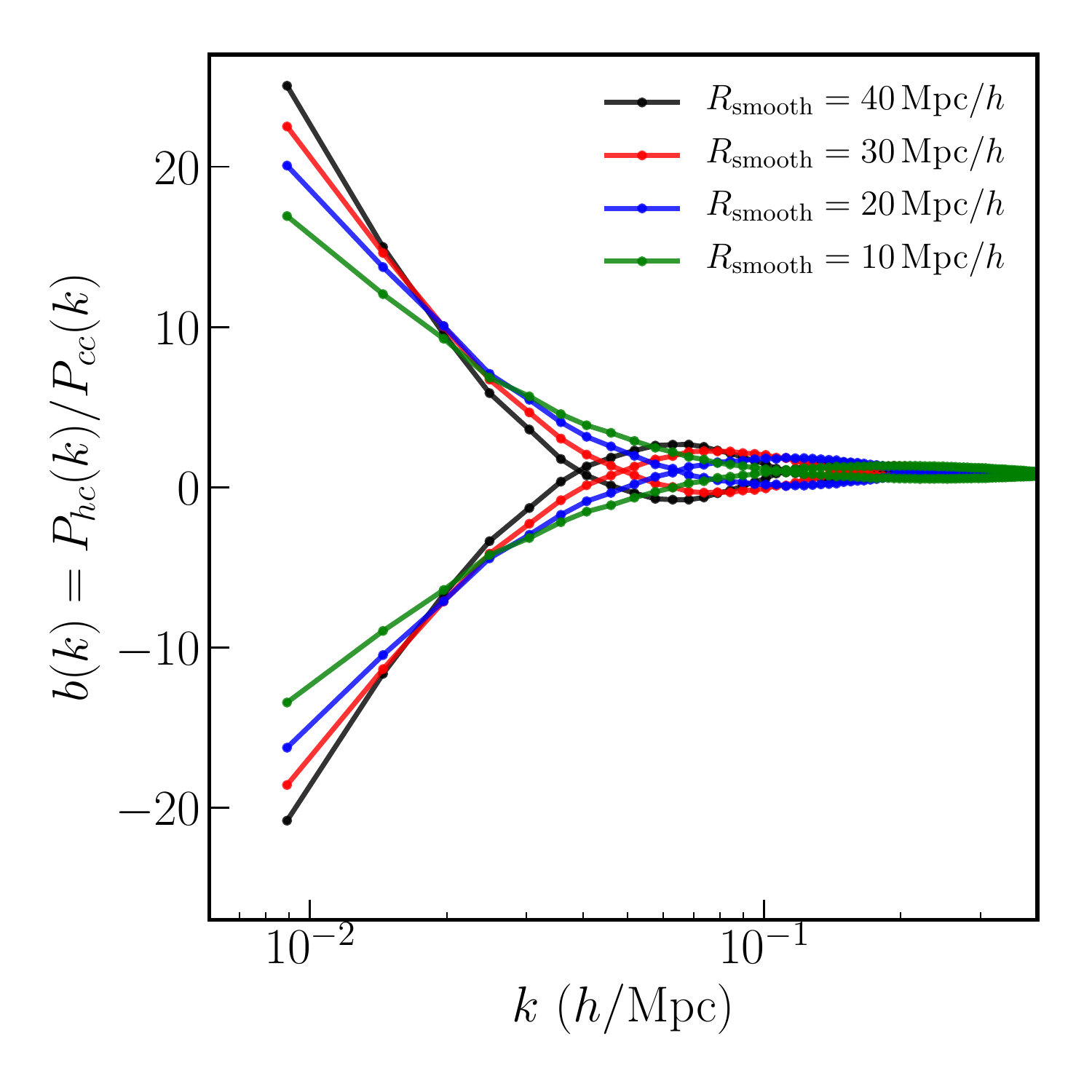}
\includegraphics[width=0.4\textwidth]{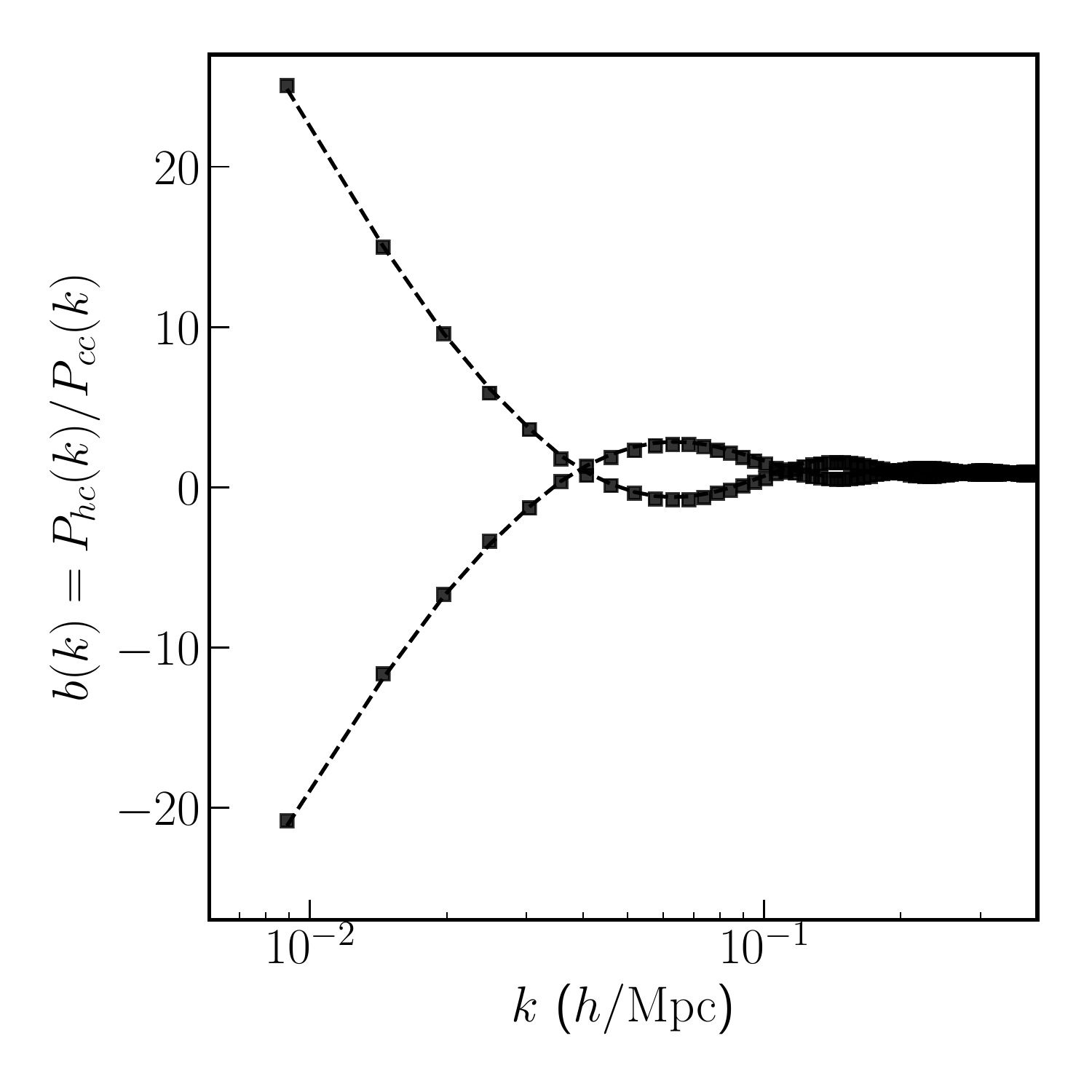}
\caption{\emph{Left panel:} Bias of the two halo populations when split on the quantity $\delta_m - \langle \delta_m \mid \delta_c\rangle$ (see text for details) for different smoothing scales $R_{\rm smooth}$, and for $M_\nu = 0.10\,$eV. \emph{Right panel:} For smoothing scale of $40\,$Mpc$/h$, we plot the data points using the black squares, and the best fit curve from Eq. \ref{eq:fit_equation} with the black dashed line.}
\label{fig:final_split}
\end{figure}

\subsection{Combining CDM and total matter environments}

While the split on the large scale neutrino overdensity produces a strongly scale-dependent linear bias, there is no feasible way of measuring the neutrino density field directly in data. Therefore, we try to reproduce this strong scale-dependent bias in the simulations using information about only the CDM ($\delta_c$) and matter overdensities ($\delta_m$), which would, in principle, be detectable in the data from various cosmological surveys (CDM from galaxy clustering and total matter from weak lensing). First, we re-do the above analysis using the value of $\delta_m - \delta_c$ to split the halos into two sets. In practice, we smooth the two overdensity fields individually on each smoothing scales before taking the difference. Notice that we have not assumed anything directly about neutrinos in this environment definition. The results of this analysis are plotted in Fig. \ref{fig:cdm_minus_matter}. Also in this case, we find that as the smoothing scale increases, the difference in the bias of the two sets increases, but more interestingly, the bias is no longer scale independent on large scales. The departure from scale independence is stronger for the larger smoothing scales.
The scale dependence in Fig. \ref{fig:cdm_minus_matter} is different from the one in Fig. \ref{fig:nu_split}. Whereas the latter was proportional to $P_{c\nu}/P_{cc}(k)$, the former is proportional to $P_{c\nu}/P_{cc}(k)-1$, and the bias indeed starts to approach the value of the full sample at large scale. 
This is a non trivial check of the the model discussed in Sec. \ref{sec:analytical}, which is able to capture the main physical effects at play in the simulation.

Even though using the combination $\delta_m - \delta_c$ as the environment defintion to split halos shows some scale-dependence in the linear bias, the effect is nowhere as strong as when $\delta_\nu$ is used to define the environment. We show below that we can construct a different combination of large scale CDM and matter environments which can produce similar scale-dependent linear bias as in the case of splitting on $\delta_\nu$. 
In particular, we tabulate the value of the smoothed $\delta_{m,i}$ and $\delta_{{\rm CDM}, i}$ at the position of every halo (labeled by $i$), and compute the following quantity for every halo:
\begin{equation}
\delta_{d,i} = \delta_{m,i} - \langle \delta_{m,i} \mid \delta_{c,i}\rangle \, 
\label{eq:difference_field}
\end{equation}
To compute the average relation between $\delta_{m,i}$ and $\delta_{{\rm CDM},i}$ on the RHS of Eq. \ref{eq:difference_field}, we simply fit a straight line through all the points in the $\delta_m$, $\delta_c$ plane. We have checked that using a higher order fitting scheme to define the average relationship between $\delta_{m,i}$ and $\delta_{c,i}$ does not change our results for the range of smoothing scales considered here.

Once we have the list of $\delta_{d,i}$, we split the full halo sample into two, one for which $\delta_{d,i}>0$ and the other for which $\delta_{d,i}<0$. Note that while this choice of splitting the halos does not produce exactly the same number of halos in each population, the difference in the number of halos is small compared to the total number of halos. We then proceed as earlier,  to compute the bias of each population, and plot the results in Fig. \ref{fig:final_split}. On large scales, the bias of the populations once again show a very strong scale dependence (similar to the one we got when splitting on neutrino overdensity environment). As before, the shape of the linear scale-dependent bias is set by the relative shapes of the neutrino and CDM transfer functions, and proportional to $P_{c\nu}/P_{cc}$. We emphasize once again that this strong departure from scale independent bias on large scales shows up without using any explicit information about the neutrino overdensity fields.
We plot the results of the best fit using this functional form in \ref{eq:fit_equation} for the split using $R_{\rm smooth}=40\,$Mpc$/h$ in the right hand panel of Fig. \ref{fig:final_split}. We see that the simple functional form of Eq. \ref{eq:fit_equation} is able to fully capture the scale dependence.

\begin{figure}
\centering
\includegraphics[width=0.4\textwidth]{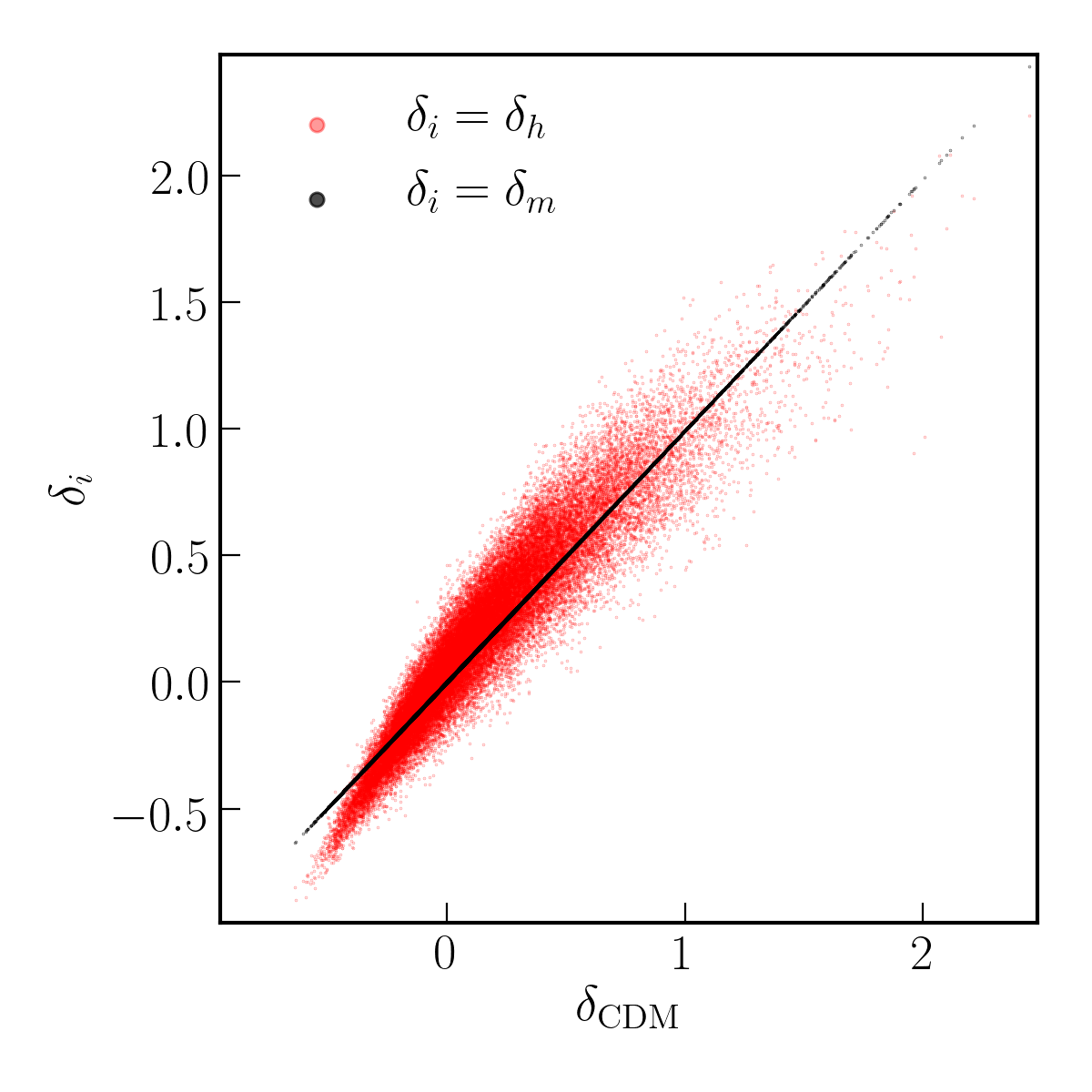}
\includegraphics[width=0.4\textwidth]{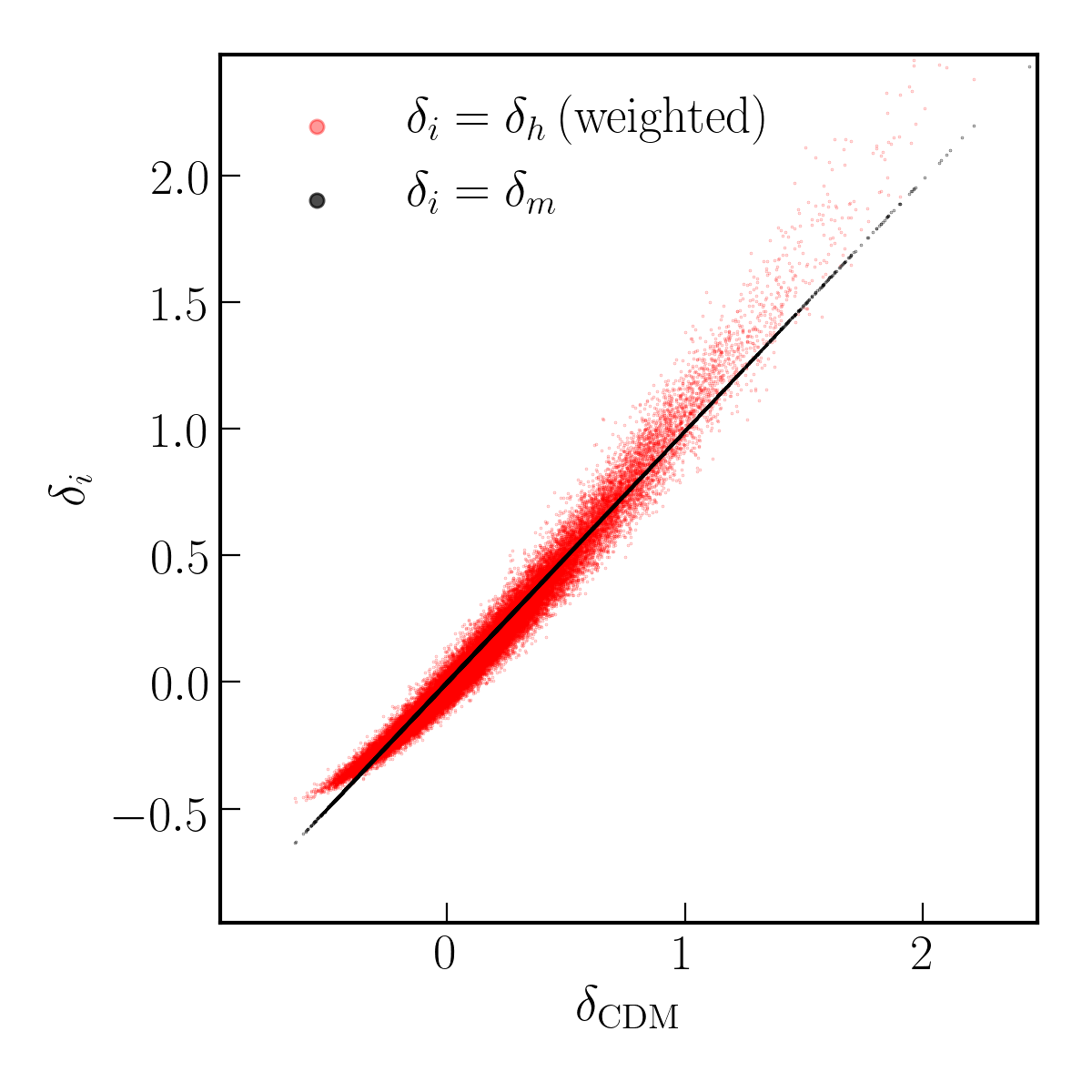}
\caption{\emph{Left panel:} Scatter in the $\delta_h-\delta_c$ relation compared to the much tighter correlation between $\delta_m-\delta_c$ where all quantities are smoothed on scale $20\,$Mpc$/h$. This scatter introduces extra noise when trying to split quantities using the value of $\delta_h$. \emph{Right panel:} Same as the left panel, except that $\delta_h$ comes from the mass weighted halo overdensity field. The scatter in the relation is reduced when the halos are mass-weighted, but the scatter is still larger than the spread in $\delta_c-\delta_m$. Note that we have rescaled all $\delta_h$ on the right panel to match the overall slopes.}
\label{fig:scatter}
\end{figure}

\begin{figure}
\centering
\includegraphics[width=0.4\textwidth]{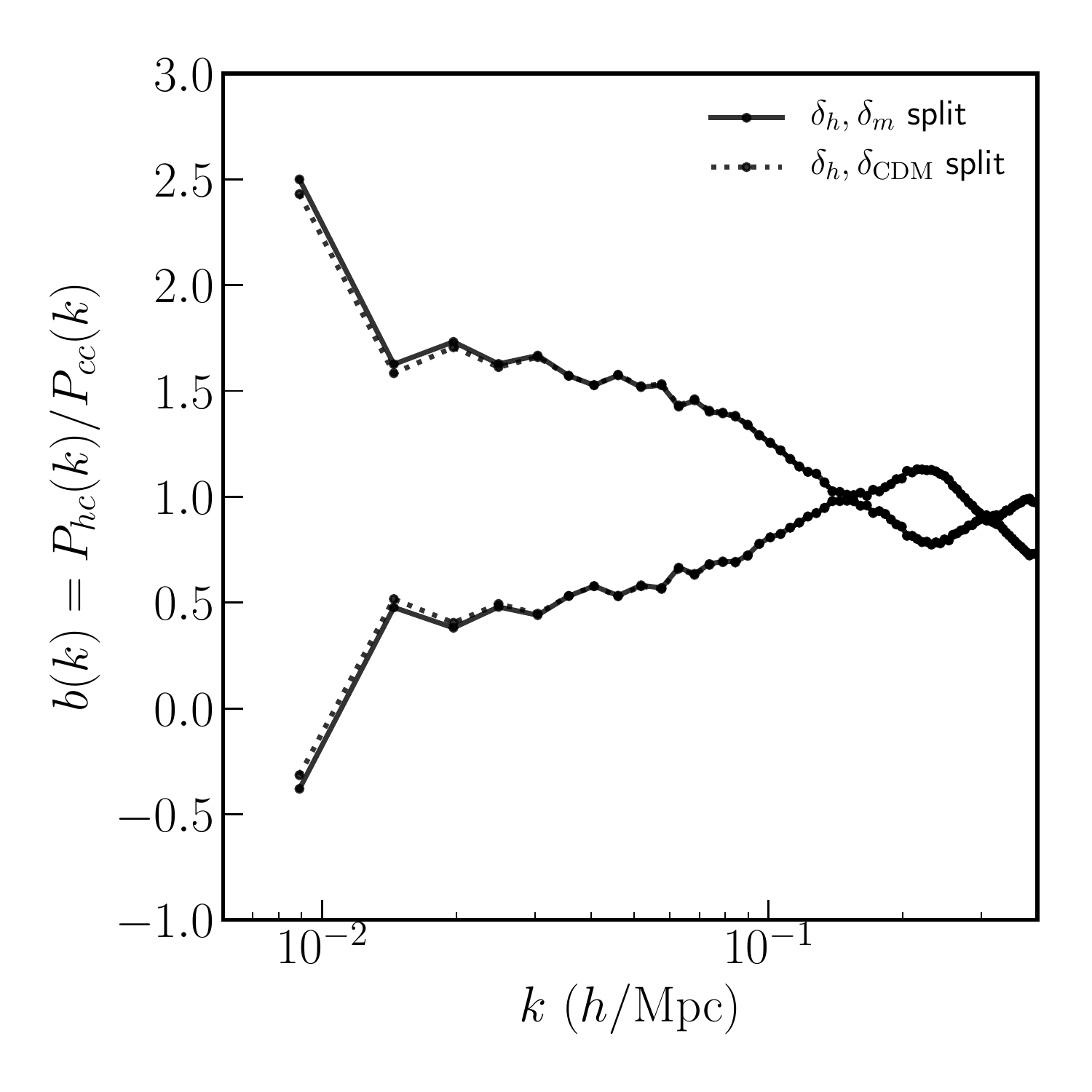}
\includegraphics[width=0.4\textwidth]{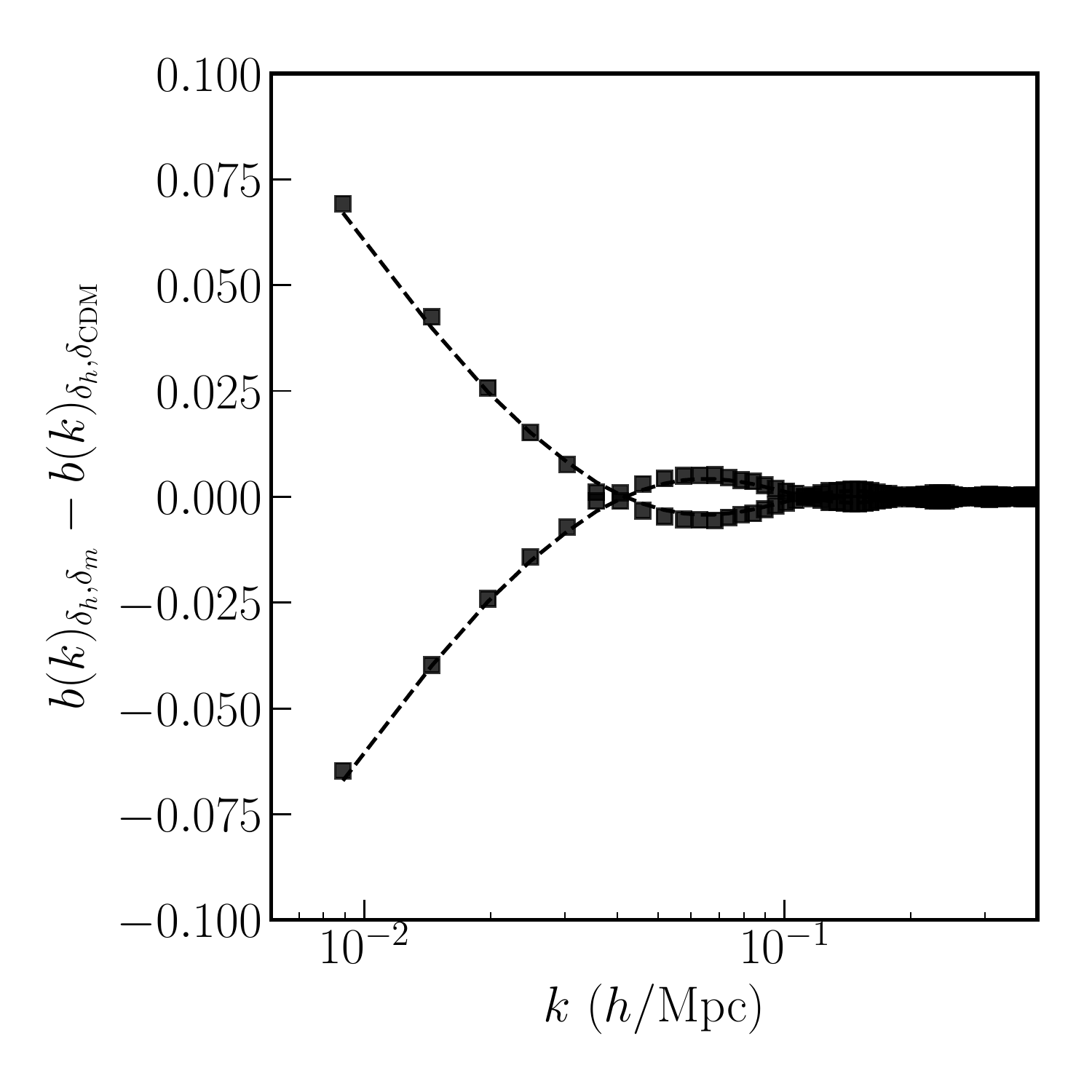}
\caption{On the left panel, we plot the bias of the halo populations split on $\delta_m - \langle \delta_m\mid \delta_h\rangle$ for a smoothing scale $R_{\rm smooth}=40\,$Mpc/h, and $M_\nu=0.10\,$eV with the solid black curves. The dotted black curves represent the bias of the populations when split on $\delta_c - \langle \delta_c\mid \delta_h\rangle$. The right hand panel shows the difference between the solid and dotted curves on from the left panel using the black squares. The dashed lines show the best fit using Eq. \ref{eq:fit_equation}.}
\label{fig:unweighted_halos}
\end{figure}

\subsection{Using biased tracers to define environment}
\label{sec:stochasticity}
So far, we have outlined a new procedure that maximizes neutrino masses effects on halo clustering. However, we have not considered any source of noise in the density fields that are used as inputs to our calculations. 
We now explore one of the sources of noise that is expected in real data. In cosmological surveys, we do not have direct information about the CDM field, but instead, have to infer it from the clustering of biased tracers of this field, such as dark matter halos, or galaxies. To quantify the effect of this on the behavior of the bias on large scales, we now use the overdensity field of halos $\delta_h$ instead of the CDM field $\delta_c$. Note that, on large scales, the halo field is related to the underlying CDM field through a deterministic bias term, but also includes stochasticity related to the finite number of tracers:
\begin{equation}
\delta_h (k) = b(k)\delta_c (k) + \epsilon(k) \,,
\label{eq:halo_bias_linear}
\end{equation}
where $b(k)$ is scale independent on large scales, and $\epsilon$ represents the stochasticity. If, at a given scale, the stochasticity is much larger than the size of the effect produced by neutrinos, the signal that we see will be washed away. We illustrate this in the left panel of Fig. \ref{fig:scatter} where the spread in the $\delta_c-\delta_h$ relation is much wider than the spread in the $\delta_c-\delta_m$ relation, even when quantities are smoothed on $20\,$Mpc$/h$. Therefore, we expect much of our signal to be washed out by the stachatic term of Eq. \ref{eq:halo_bias_linear}.

We now compute $\delta_{d,i} = \delta_{m,i}-\langle \delta_{m,i}\mid \delta_{h,i} \rangle$, where, once again, we perform a linear fit of $\delta_{m,i}$ and $\delta_{h,i}$ to obtain the mean relationship between the two. Note that this procedure accounts for the constant linear bias term in Eq. \ref{eq:halo_bias_linear}. As before, we split the full halo sample into two based on whether $\delta_{d,i}$ for halo $i$ is positive or negative. We plot the results for the bias using the solid black lines on the left panel of Fig. \ref{fig:unweighted_halos}. The smoothing scale used was $40\,$Mpc/h. 

To better understand the contribution of stochasticity to the behavior of the bias of the two populations, we also split the halos using $\delta_c$, instead of $\delta_m$, i.e. by defining $\delta_{d,i} = \delta_{c,i}-\langle \delta_{c,i}\mid \delta_{h,i} \rangle$. Note that since we assumed Eq. \ref{eq:halo_bias_linear} to be true, this way of splitting the halos should not have any information about the neutrinos. The results of this split are plotted in the dotted curves in the left panel of Fig. \ref{fig:unweighted_halos}.
By comparing the solid and dotted lines in Fig. \ref{eq:difference_field}, we see that, as expected from the large scatter in the left panel of Fig. \ref{fig:scatter}, the extremely strong scale dependence from neutrinos that we found previously is mostly washed away due to the stochasticity in the halo field. Further, the difference in the large scale behavior of the populations split using $\delta_m\,,\delta_h$ is very similar to the behavior of the subsamples from the $\delta_c\,,\delta_h$.  However, we find that the full signal is not washed away. Instead, when we compute the bias  using the $\delta_m$, $\delta_h$ split minus the the bias using the $\delta_c$, $\delta_h$ split, we find that the difference still scales exactly with $P_{c\nu}/P_{cc}$, but the amplitude of the signal is much smaller, by approximately a factor of 100, than when using the true CDM field. This is plotted in the right hand panel of Fig. \ref{fig:unweighted_halos}.

\begin{figure}
\centering
\includegraphics[width=0.4\textwidth]{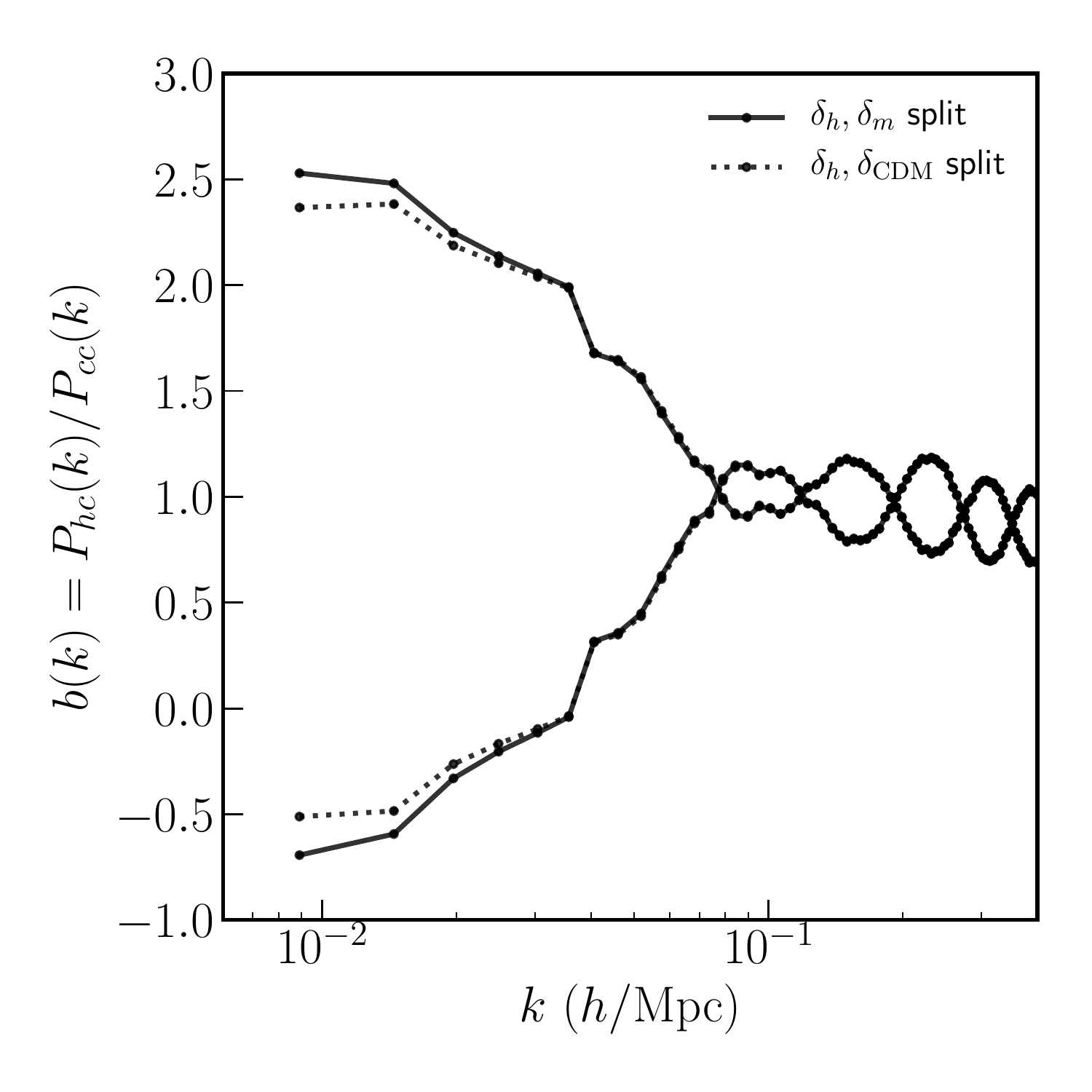}
\includegraphics[width=0.4\textwidth]{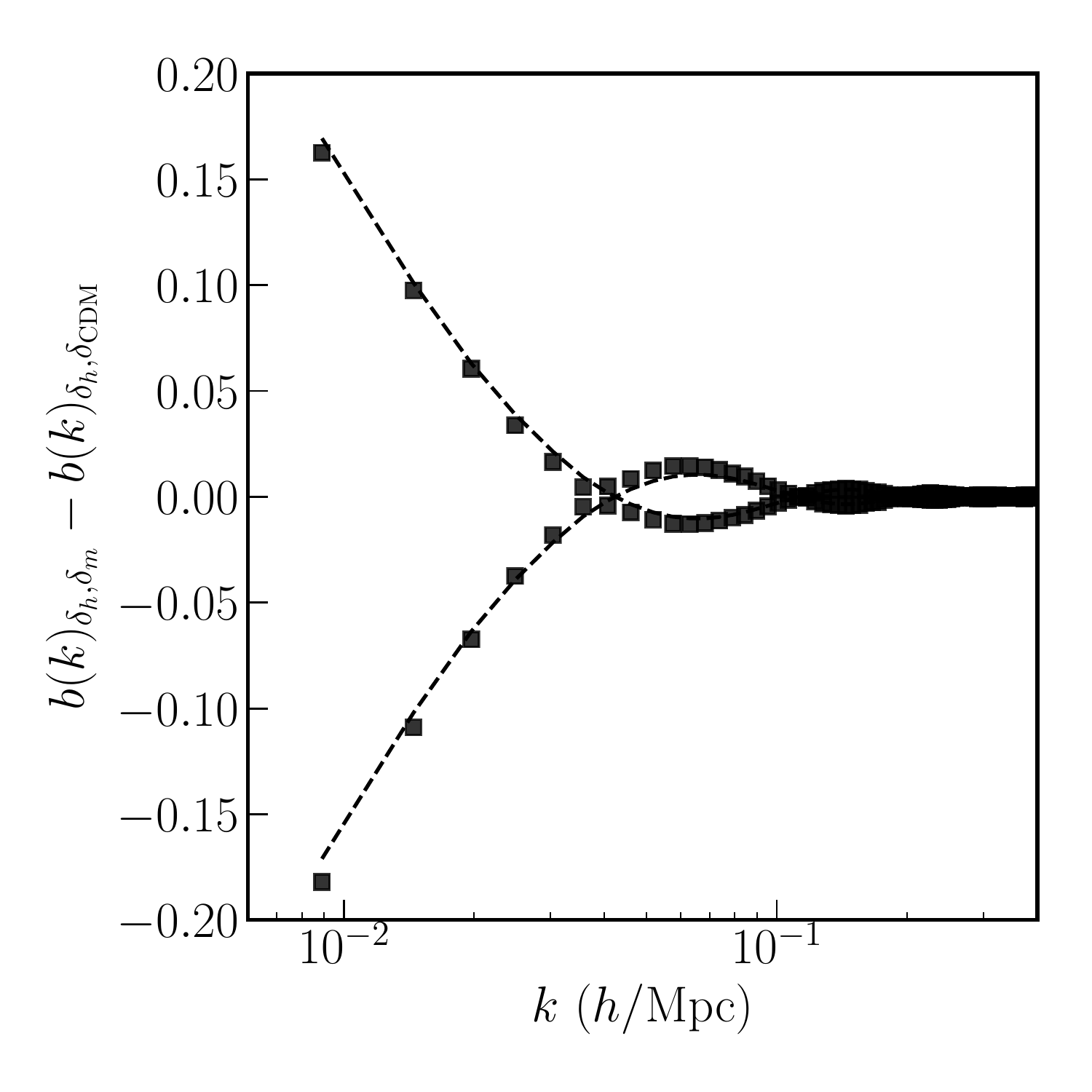}
\caption{Same as Fig. \ref{fig:unweighted_halos}, but using the mass weighted halo field while determining the split into two populations. For the right hand panel, we plot the data points with the solid black squares, and the best fit curve from Eq. \ref{eq:fit_equation} fitted to the difference of the two biases using the dashed black line.}
\label{fig:weighted_halos}
\end{figure}

Since the stochastic part in Eq.  \ref{eq:halo_bias_linear} is responsible for reducing the strength of the scale-dependent bias signal, we explore how, for a fixed number density, we can try to reduce the stochasticity in the halo overdensity field. Previous studies \cite{Cai:2010gz,Seljak:2009af,PhysRevD.82.043515} have shown that the stochasticity in the halo density field can be reduced by mass-weighting the halos while computing the density. We implement this in our analysis by weighting every halo by its mass during the CIC deposition step while computing the halo overdensity field, $\delta_h$. The reduction in the scatter in the values of the smoothed $\delta_c\,,\delta_h$ at $20\,$Mpc$/h$ for each halo is illustrated in the right panel of Fig. \ref{fig:scatter}. We note that the scatter is still larger than the spread in the $\delta_c\,,\delta_m$ relation, where the signal comes from.

The rest of the analysis is carried out as before. The results are shown in Fig. \ref{fig:weighted_halos}. We find that the mass weighting indeed helps increasing the difference between the populations identified by the $\delta_m\,,\delta_h$ and $\delta_c\,,\delta_h$ splits. On the right hand panel of the figure, we plot the data points of the difference in bias using the solid squares, and the best fit curve using Eq. \ref{eq:fit_equation} with the dashed black line. This demonstrates how the shape of the bias difference from the two splits is well described by the shape of the neutrino transfer functions. However, as the left hand panel of Fig. \ref{fig:weighted_halos} demonstrates, mass weighting the halos is still not enough to clearly show the scale-dependent bias effect in only the $\delta_m\,,\delta_h$ splitting scheme. This suggests that while the shape of the halo bias on large scales contains information on neutrino masses, its main broadband shape is dominated by CDM environmental effects, that will downgrade the signal-to-noise ratio of this measurement to neutrino masses. This means that more sophisticated methods may be needed to reconstruct the true underlying CDM density field from the observed galaxy fields in cosmological surveys (c.f. \cite{Modi:2018cfi}) in order for the environmental scale-dependent bias effect to be clearly detected. 

\begin{figure}
\centering
\includegraphics[width=0.4\textwidth]{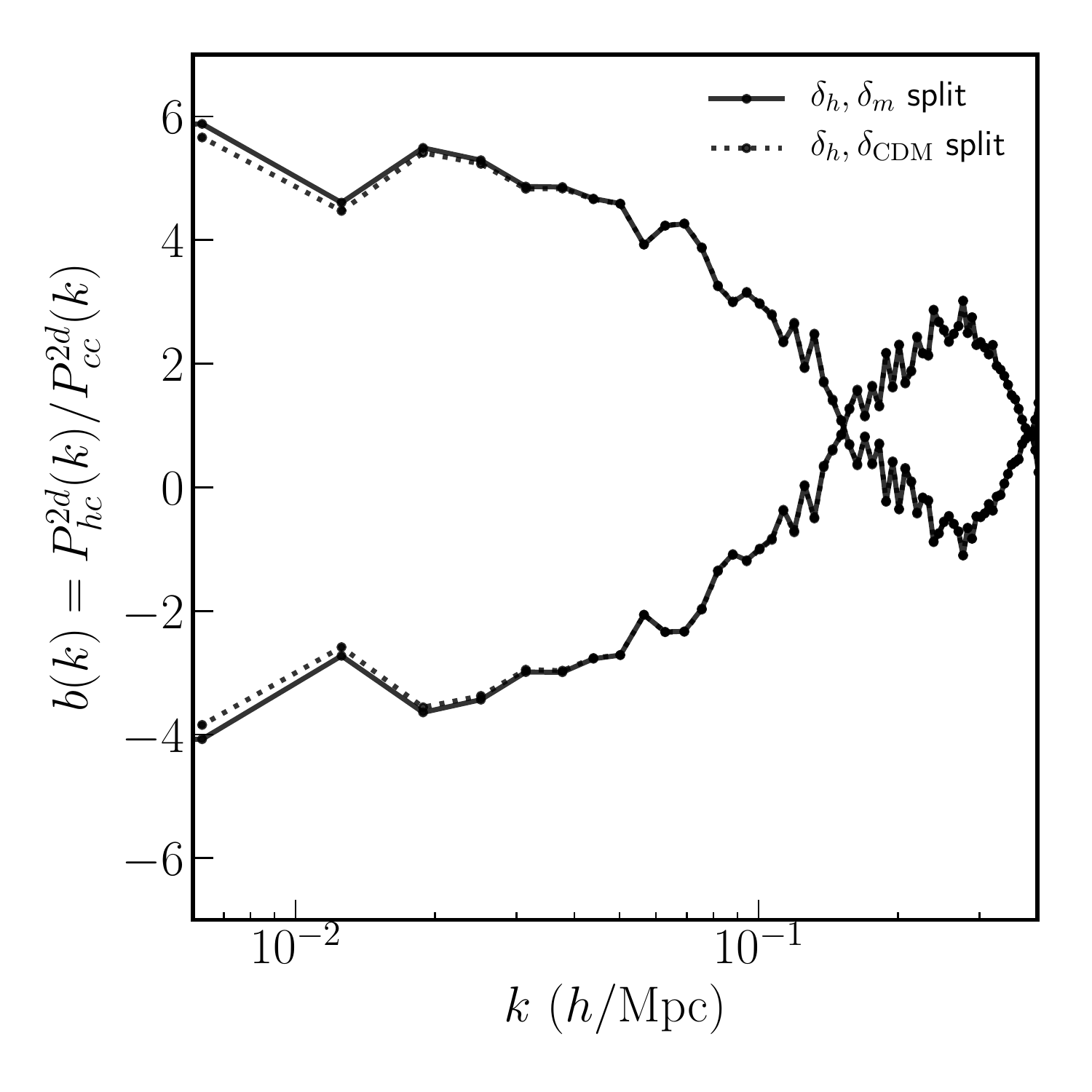}
\includegraphics[width=0.4\textwidth]{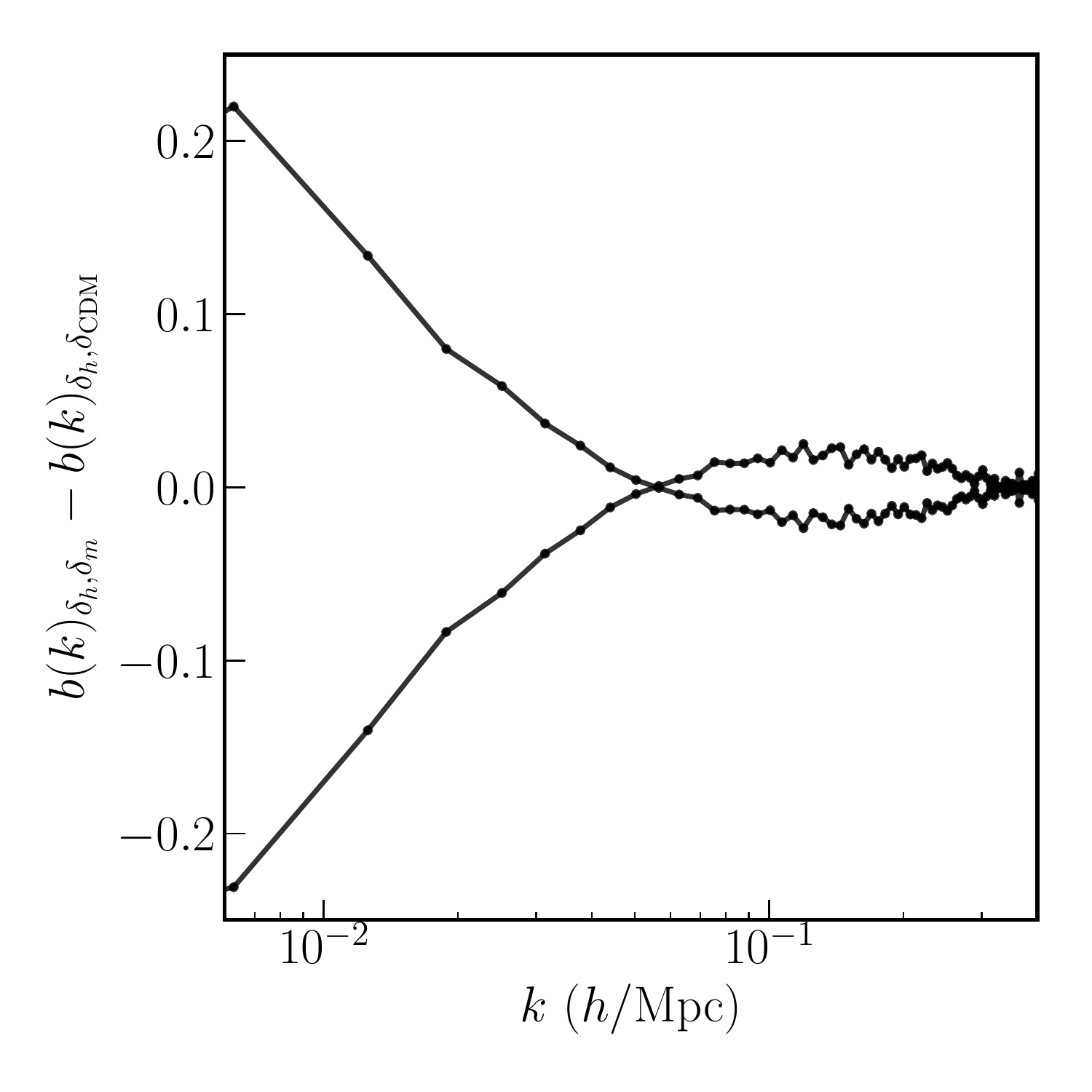}
\caption{Same as Fig. \ref{fig:weighted_halos}, but with all quantities projected along one of the directions in the simulation box. The smoothing scale used is $10\,$Mpc/$h$, and with $M_\nu=0.10\,$eV. Projecting to 2d does not significantly change the behavior of the large scale bias.}
\label{fig:2d_weighted_halos}
\end{figure}

\subsection{Environmental bias in projected fields}
Finally, we explore how the large scale bias of different populations of halos in the simulation behave when we look at all quantities in projection, rather than assuming we have the full 3-dimensional fields. This is especially relevant in the context of lensing, where the measured shear or convergence are quantities which are directly related to a weighted projection of the underlying 3-dimensional total matter field. As a crude approximation of this, we repeat our analysis using only quantities that have been projected along one of the simulation directions. For example, the halo overdensities are calculated by throwing away information about the $z$ position of the halos, and depositing onto a 2-dimensional grid in the $x-y$ plane. A similar procedure is adopted to get the CDM and matter density fields in 2 dimensions. We note that in this case, the smoothing on scale $R_{\rm smooth}$ is done on the projected fields, i.e. we use a 2-dimensional top-hat aperture for the smoothing procedure. 

We specifically consider the case where we perform the splits using $\delta_m$ and the mass weighted version of $\delta_h$.  The smoothing radius we have used in this specific example is $10\,$Mpc/$h$. We present the results in Fig. \ref{fig:2d_weighted_halos}. We find that, qualitatively, the results remain the same as those in Section \ref{sec:stochasticity}. Specifically, we find that when using the halo density field instead of the CDM field, the stochasticity washes out most of the signal of scale-dependent bias coming from the presence of massive neutrinos when using $\delta_m$ and $\delta_h$ split. Similar to our results in 3 dimensions, this is true even for the mass weighted field, as seen in Fig. \ref{fig:2d_weighted_halos}. However, as in the 3-dimensional case, the linear scale-dependent bias is clearly visible in the difference in the bias of the halo populations created by taking the $\delta_m\,,\delta_h$ and $\delta_c\,,\delta_h$ splits.

\section{Summary and Discussion}
\label{sec:conclusions}

In the standard $\Lambda$CDM cosmology, the large scale bias of nonlinear objects is expected to be scale-independent. The presence of both non-Gaussianities and massive neutrinos can lead to scale-dependent bias, a unique signature that can be used to constrain them.
In the case of massive neutrinos, the shape of this scale dependence is set by the relative difference between the transfer function of neutrinos and CDM. Since the transfer function of neutrinos is determined by the neutrino mass, a measurement of departure from scale independent bias on linear scales can be converted into constraints on the neutrino mass. While this scale-dependent bias is a unique signature of massive neutrinos on large scale structure, the size of the signal is usually very small, of the order of $f_\nu$.

In this paper, we have shown for the first time, how large departures from scale independence bias on large scales can be achieved by using information about the local environment of dark matter halos. For the purpose of this analysis, the local environment is defined by the value of the either the CDM, neutrino, or total matter density field smoothed on some scale; we have considered scales from $10\,$Mpc$/h$ to $40\,$Mpc$/h$. We have shown that if we split the halo population in the simulation volume into two populations, based on the enclosed neutrino overdensity around each halo, we obtain a very strong scale-dependent bias on large scales. The amplitude of the scale-dependence is much higher than the expected scaling with $f_\nu$, and increases when the smoothing scale used to define the environment rises. In contrast, the bias on large scales always remains scale independent when we perform the same split but using the enclosed CDM overdensity on the defined smoothing scales. This suggests that this effect is entirely due to the presence of massive neutrinos; this is the reason why the signal-to-noise ratio of this statistics is much larger than traditional observables.

Since for realistic cosmologies, $f_\nu \lesssim 1\%$, the ability to boost the neutrino signal using this method could be crucial in the eventual detection of the neutrino mass from Large Scale Structure observations. Besides, since this effect takes place on very large scales, it will be less affected by traditional complications such as redshift-space distortions, non-linearities and baryonic effects. While the actual values of the bias change as a function of the scale used to define the environment, we have shown that for all smoothing scales, and for all masses, the shape of the scale-dependent bias is extremely well fit by the ratio of the neutrino and CDM transfer functions, illustrating that this effect is unique to massive neutrino cosmologies. Since the neutrino transfer functions are identical to the CDM transfer functions on scales much larger than the free streaming length, the bias returns to being scale independent on the very largest scales ($\gtrsim 3{\rm Gpc}/h$). Thus, this effect will not be degenerate with the scale-dependent bias that primordial non-Gaussiantities models induce, where the bias is expected to scale as $k^{-2}$ for small $k$. Our method allows us to constrain neutrino masses by measuring the shape of the halo bias.

Since the neutrino density field itself is not directly observable, we have investigated whether the large neutrino environmental scale-dependence
can be reproduced using only information about the large-scale CDM and total matter fields. In principle, one can use the galaxy positions to infer the underlying CDM density field using a linear bias model. Similarly, since photon deflection depends on the gravitational potential, sourced by all matter, weak lensing measurements can be used to infer the distribution of total matter on large scales. We have shown that if we split halos according to a combination of their $\delta_m$ and $\delta_c$ environment, we observe a very similar scale-dependence bias than the one we get by splitting from neutrino overdensity.

Rather than true underlying CDM distribution $\delta_c$, a closer analog to what is truly measured in galaxy surveys is the halo overdensity field $\delta_h$. Halos are biased tracers of the CDM field, and the relationship is expected to be linear on large scales. However, there is also a stochastic contribution to the halo field which is uncorrelated with the CDM field. This uncorrelated piece acts as a source of noise for the signal that we are looking for. We find that when halos are split according to a combination of $\delta_m$ and $\delta_h$ environments, instead of $\delta_m$, $\delta_c$ environments,
most of our signal vanishes due to the stochasticity. However, a part of the signal still remains, and its amplitude is still larger than $f_\nu$. We have shown that the amplitude of the neutrino signal increases when stochasticity in the halo density field is reduced by mass-weighing the halos. A similar noise reduction in the definition of the environment in galaxy surveys can be achieved by assuming a relationship between halo mass and galaxy luminosity and weighing the galaxies by their luminosity. However, we possibly need more sophisticated methods of determining the true underlying CDM distribution from the galaxy distribution to achieve the levels of noise needed to cleanly recover the environmental neutrino scale-dependent bias signal from upcoming surveys.

The weak lensing measurements from cosmological surveys do not measure the 3-dimensional distribution of matter, but rather, a weighted projection of it along the line of sight. Further, in photometric surveys, lack of precise redshift information of galaxies implies that, often, it is only possible to define the galaxy density field in broad redshift bins. To roughly approximate these effects, we have shown that the results obtained using the 3-dimensional field also hold when all the fields are projected down to 2 dimensions. 

In this paper we have proposed a new method to weigh neutrinos by detecting their effects on the large-scale halo/galaxy bias. Our methods magnifies the signal-to-noise ratio of neutrino masses effects on large-scale structure by exhibitng a large scale-dependence halo/galaxy bias on linear scales. 
The departure from scale-independence can be much larger than $f_\nu$, and increases with lower neutrino masses. On the other hand, the effects of cosmic variance are much larger on these scales. Therefore, as part of future work, we plan to quantify the actual constraints on the neutrino mass that can obtained using this method in future cosmological surveys like LSST, and Euclid.

\acknowledgments
We thank Ravi Sheth, Aseem Paranjape, Cristino Porciani, David Spergel and Tom Abel for useful conversations. This work used the Sherlock cluster at the Stanford Research Computing Center. Simulations were run on the Rusty cluster at the Center for Computational Astrophysics and the Gordon cluster at the San Diego Supercomputer Center. The \textsc{HADES} and \textsc{Quijote} simulations are publicly available at \url{https://franciscovillaescusa.github.io/hades.html} and \url{https://github.com/franciscovillaescusa/Quijote-simulations}, respectively. This work was supported by collaborative visits funded by the Cosmology and Astroparticle Student and Postdoc Exchange Network (CASPEN). The work of FVN is supported by the Simons Foundation. MV is supported by INDARK PD51 INFN grant.

\bibliography{ref}
\bibliographystyle{JHEP}
\end{document}